\documentclass[journal]{IEEEtran}
\usepackage{cite}
\usepackage{graphicx}
\usepackage{algorithm,algpseudocode}
\usepackage{algorithmicx}
\usepackage{subfig}

\usepackage{amssymb}
\usepackage{mathrsfs}
\usepackage{amsmath}

\usepackage[sans]{dsfont}
\usepackage{stmaryrd}
\usepackage{pifont}
\usepackage{wasysym}

\usepackage{array}
\usepackage{url}
\usepackage[usenames]{color}

\algnewcommand{\LeftComment}[1]{\Statex \(\triangleright\) #1}

\newtheorem{innercustomthm}{Theorem}
\newenvironment{customthm}[1]
  {\renewcommand\theinnercustomthm{#1}\innercustomthm}
  {\endinnercustomthm}

\newtheorem{definition}{Definition}

\newtheorem{lemma}{Lemma}

\begin{document}

\title{Cost Minimizing Online Algorithms for Energy Storage Management with Worst-case Guarantee}

\author{
Chi-Kin~Chau, {\em Member}, {\em IEEE}, Guanglin~Zhang,  {\em Member}, {\em IEEE}, and Minghua~Chen, {\em Senior Member}, {\em IEEE}
\thanks{C.-K. Chau is with Department of EECS at Masdar Institute of Science and Technology, UAE (e-mail: ckchau@masdar.ac.ae).}
\thanks{G. Zhang is with Department of Communication and Electronic Engineering at Donghua University, China (e-mail: glzhang@dhu.edu.cn). G. Zhang was supported by the NSF of China (Project No. 61301118), the Innovation Program of Shanghai Municipal Education Commission (Project No. 14YZ130) and the International S\&T Cooperation Program of Shanghai Science and Technology Commission (Project No. 15220710600).}
\thanks{M. Chen is with Department of Information Engineering at the Chinese University of Hong Kong, Hong Kong (e-mail: minghua@ie.cuhk.edu.hk). M. Chen was supported by National Basic Research Program of China (Project No. 2013CB336700) and the University Grants Committee of the Hong Kong Special Administrative Region, China (Theme-based Research Scheme Project No. T23-407/13-N and General Research Fund No. 14201014).}
}

\pagestyle{plain}
\thispagestyle{plain}

\maketitle

\begin{abstract}
The fluctuations of electricity prices in demand response schemes and intermittency of renewable energy supplies necessitate the adoption of energy storage in microgrids. However, it is challenging to design effective real-time energy storage management strategies that can deliver assured optimality, without being hampered by the uncertainty of volatile electricity prices and renewable energy supplies. This paper presents a simple effective online algorithm for the charging and discharging decisions of energy storage that minimizes the electricity cost in the presence of electricity price fluctuations and renewable energy supplies, without relying on the future information of prices, demands or renewable energy supplies. The proposed algorithm is supported by a near-best worst-case guarantee (i.e., competitive ratio), as compared to the offline optimal decisions based on full future information. Furthermore, the algorithm can be adapted to take advantage of limited future information, if available. By simulations on real-world data, it is observed that the proposed algorithms can achieve satisfactory outcomes in practice.
\end{abstract}


\section*{Nomenclature}
\addcontentsline{toc}{section}{Nomenclature}
\begin{IEEEdescription}[\IEEEusemathlabelsep\IEEEsetlabelwidth{${\cal N}$}]

\item[$t$] Index of an one-hour time slot 

\item[$B$] Capacity of energy storage

\item[$p(t)$] Electricity market price at time slot $t$, where\\ $m \leq p(t)\leq M$

\item[$\varphi$] Ratio between maximum and minimum prices $\big(\frac{M}{m}\big)$

\item[$a(t)$] Excessive demand at time slot $t$

\item[$\max(a)$] \quad\ Peak excessive demand $(\max_t a(t))$

\item[$x(t)$] Energy storage level at time slot $t$

\item[$\eta_{\rm c}(\le 1)$] \quad\ Charging efficiency of energy storage

\item[$\eta_{\rm d}(\ge 1)$] \quad\ Discharging efficiency of energy storage

\item[$\mu_{\rm c}$] Charging rate constraint

\item[$\mu_{\rm d}$] Discharging rate constraint

\item[$d(t)$] Energy discharged from energy storage at time slot $t$

\item[$r(t)$] Excessive renewable energy at time slot $t$

\item[$r_{\rm b}(t)$] Energy from renewable energy to energy storage at time slot $t$

\item[$v(t)$] Energy from the grid at time slot $t$

\item[$v_{\rm a}(t)$] Energy from the grid for demand at time slot $t$

\item[$v_{\rm b}(t)$] Energy from the grid to energy storage at time slot $t$

\item[$\rho$] Ratio of total excessive renewable energy over total excessive demand, normalized by charging and discharging efficiencies $\big(\frac{\eta_{\rm c}}{\eta_{\rm d}} \cdot \frac{\sum_{t=1}^T r(t)}{\sum_{t=1}^T a(t)}\big)$

\item[$\theta$] Threshold of market price that triggers charging operation in the threshold based algorithm

\item[$\hat{B}$] Maximum energy storage level, up to which the threshold based algorithm charges from the grid

\item[$W$] Lookahead window size for the lookahead algorithm

\item[$\lbrack\ \cdot\ \rbrack^+$] Positivity operator ($\max\{\ \cdot\ , 0 \}$)

\end{IEEEdescription}

\section{Introduction} \label{sec:intro}

The rise of renewable energy can deliver a clean energy future, while innovative demand response management by time-varying dynamic pricing schemes can facilitate more efficient balance between energy supplies and demands. Microgrids are a new paradigm of electric systems that integrate both local renewable energy supplies and distributed power management. Thus, energy storage is crucial in microgrids for harnessing excessive renewable energy and reducing the electricity cost in a fluctuating electricity market.

To harness both renewable energy and dynamic electricity prices, energy storage systems are required to make real-time charging and discharging decisions intelligently. However, the energy storage management decisions for dynamic electricity prices and renewable energy are coupled with one another. The decisions favoring electricity acquisition from the grid may saturate the energy storage too soon, preventing it from harnessing excessive renewable energy in the future. On the contrary, the decisions that reserves more capacity for storing future renewable energy may limit the chance of acquiring electricity at low electricity prices. Hence, it is important to strike a subtle balance between the two factors. Nonetheless, there is considerable uncertainty in both renewable energy availability and electricity price fluctuations. The fluctuations of electricity prices tend to reflect the wide-area demands, whereas the renewable energy availability may only be specific to the local regions of microgrids. The different considerations of uncertainty present a major challenge to the design of an effective strategy for energy storage management.

In the extant literature, there are three common approaches for energy storage management. First, one can rely on predictions on the renewable energy availability, price fluctuations and demands \cite{Fan12}, which require accurate modeling and extensive training data. The drawback is the heavy reliance on accurate prediction models or specifically trained learning classifiers for the particular environments, which are difficult to be adopted to new environments with noisy or limited historical data for calibration.

Second, one can utilize stochastic optimization \cite{Costa07, VHMS13}, which relies on standard probability models to handle uncertainty or noisy data \cite{Gamal11}. The solutions are usually obtained in the sense of probabilistic expectation. There may be considerable deviation between the real-world outcomes and the standard probability model. Recently, a Lyapunov stochastic optimization approach has been proposed \cite{Neely11} for energy storage management, by which a control policy is developed that asymptotically converges to the optimal policy, when the inputs are assumed to be i.i.d. or stationary Markovian. Nonetheless, the inputs in practice (e.g., demands, renewable energy supplies, and electricity prices) are often non-stationary. Furthermore, Lyapunov stochastic optimization relies on averaging over infinite time horizon, whereas only finite time horizon is considered in practice. The asymptotic optimality result is obtained in terms of scaling-law, when the storage size is approaching infinity. However, for small-to-medium-sized energy storages, the gap between the outcome and the optimal may be large. Our experiments based on real-world traces in Sec.~\ref{sec:sim} confirm that Lyapunov stochastic optimization does not produce satisfactory outcome for small-to-medium-sized energy storages.

The third approach is based on robust optimization \cite{BertsimasS03}, which handles uncertainty by optimizing the solutions with respect to a range of parameters. To the best of our knowledge, there is no known results from robust optimization that can qualify the worst-case guarantee to the optimal solutions based on perfect knowledge without uncertainty. Also, applying the existing robust optimization to online decision-making problems over a long time horizon can suffer from the curse of dimensionality.

As a departure from the aforementioned approaches, this paper pursues an {\em online competitive algorithmic} approach, which has been employed in a wide range of online decision-making problems \cite{Borodin:1998}, without relying on the knowledge of future inputs. This approach can cope with arbitrary (non-stationary or adversarial) future inputs, with a finite or infinite time horizon, and can provide a worst-case optimality assurance in terms of competitive ratio (as compared to the offline optimal decisions based on full future information) without asymptotic or stochastic assumptions. 

We propose an effective online algorithm for energy storage management based on a simple notion of threshold based decisions, which can be conveniently adopted in the existing energy storage systems. Furthermore, we derive the lower bounds of competitive ratio for any deterministic online algorithms, which show that the proposed algorithm can attain a near-best competitive ratio. The proposed algorithm can also be adapted to take advantage of limited future information, if available. Not only supported by its worst-case guarantee, it is observed from simulations on real-world data that the proposed algorithms can achieve satisfactory outcome in practice.

The contributions of this paper are summarized as follows:
\begin{enumerate}
\item An online algorithm is proposed for tackling the electricity cost minimization problem of energy storage management with price fluctuations and renewable energy supplies. The competitive ratio for the proposed algorithms is derived against the offline optimal algorithm. 

\item The lower bounds of competitive ratio for any deterministic online algorithms are derived, which are within a constant factor of that of the proposed algorithm. 

\item An improved online algorithm is presented to take advantage of limited future information in a sliding window fashion.

\item By evaluations with real-world data traces, it is observed that the proposed algorithms achieve a satisfactory empirical optimality ratio.

\end{enumerate}

\section{Background and Related work} \label{sec:related}

There is a body of work about applying energy storage to reduce the electricity cost. For example, to design optimal scheduling algorithms for energy storage, \cite{Chandy10} formulates the problem of energy storage management by convex optimization. \cite{Gamal11} presents an optimal policy for energy storage management with fast-ramping generation. \cite{KHT11 ,VHMS13} derive energy storage control policy based on Markov decision problem. \cite{RuiZhang15} develops an online heuristic by predicting the demands in a sliding window. These results require a-priori statistical assumptions (e.g., i.i.d. prediction errors, Markovian arrivals, stationary stochastic inputs), which may not hold in practice.

The Lyapunov optimization approach \cite{Neely11} was introduced to investigate the cost minimization problem of energy storage, which assumes the inputs to be i.i.d. or Markovian random process, and relies on an asymptotic analysis by long-term averaging the total cost. Various extensions have been developed (e.g., \cite{Zhang13, Grillo12, LD14, GXB14, SDL14}). Lyapunov optimization requires stochastic assumption (e.g., stationary inputs), whereas our competitive online approach can accept arbitrary (non-stochastic) inputs. Furthermore, Lyapunov stochastic optimization relies on asymptotic analysis (i.e., infinite time horizon, and very large-sized energy storage), whereas our online algorithm can cope with finite setting, with arbitrarily sized energy storage.

Online algorithmic approach \cite{Borodin:1998} is an established approach in optimizing the performance of various systems (e.g., computer systems, microgrids \cite{Lian13}) with minimum knowledge of inputs. For an online decision problem, a sequence of inputs are revealed gradually over time. The algorithm needs to make certain decisions and generates output instantaneously over time, based on only the part of the inputs that has seen so far, without knowing the rest of the inputs in the future.  

Closely related to the results of this paper, \cite{Yaniv01} investigated the classical {\em one-way trading problem} and devised competitive online algorithms with optimal competitive ratio. In one-way trading problem, a trader needs to exchange from one currency to another currency, when given a sequence of exchange rates in an online fashion. One-way trading problem can be regarded as a continuous version to $k$-min/max search problem \cite{Yaniv01}, in which a trader is searching for the $k$-th minimum (or maximum) price of some asset, when given a sequence of prices in an online fashion \cite{lorenz2009optimal}. See the detailed definitions in the Appendix. The energy storage management problem in this paper can be regarded as a generalized one-way trading problem. However, there are surprising differences. For instance, the best possible online algorithm for one-way minimum trading problem has competitive ratio $\sqrt{M/m}$, where $M$ and $m$ are maximum and minimum market prices. In energy storage management with {\em free} renewable energy, one would think $m \to 0$, and hence the competitive ratio becomes unbounded. However, we show that this is not the case. Moreover, the energy storage management problem needs to consider additional constraints of limited capacity, charging and discharging operations.

\section{Problem Formulation} \label{sec:model}

\begin{figure}[htb!]
\centering 
\includegraphics[scale=0.3]{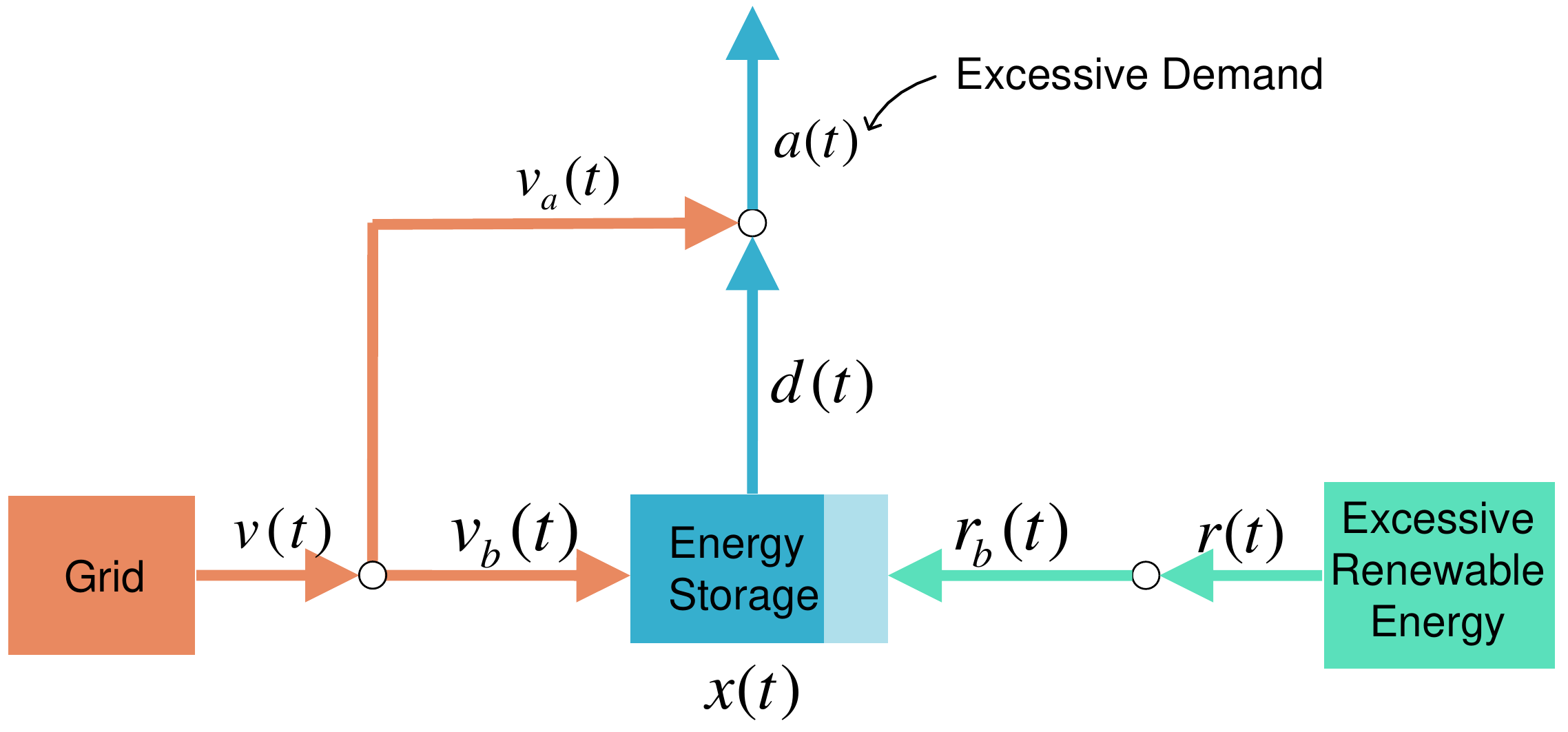} 
\caption{A depiction of the system model.} \label{fig:system} 
\end{figure}

Consider a typical scenario of an operator of a microgrid, who needs to manage different energy sources (e.g., electricity grid, energy storage and renewable energy) to minimize the total electricity cost, subject to the demand and operational constraints. The system model of such a scenario is depicted in Fig.~\ref{fig:system}, which has been widely used in the literature \cite{Neely11}. A discrete-time model is considered in this paper, such that each time slot matches the timescale at which the energy management decisions are updated. This paper assumes the duration of a time slot is one hour. We denote $t$ as a time slot index. There are totally $T$ time slots, and each has a unit length, where the inputs within one time slot are sufficiently quasi-static. For brevity, the power and energy within a time slot are referred interchangeably.

\subsection{System Model}

\noindent
The system model is consisted of the following components:
\begin{itemize}
\item
\textbf{Excessive Demand:} Consider arbitrary excessive demand (after being offset by the renewable energy at the current time slot) at time slot $t$, denoted by $a(t)$. We do not assume any specific stochastic model of $a(t)$. We simply refer $a(t)$ as the demand in this paper. 

\item
\textbf{Electricity Grid:} The system can acquire energy from the grid for the unsatisfied demand in an on-demand manner. Let the market price at time slot $t$ of the grid be $p(t)$, where $m\leq p(t)\leq M$. Denote the ratio between the maximum and minimum prices by $\varphi \triangleq \frac{M}{m}$. We do not assume any specific stochastic model on $p(t)$. Denote the acquired energy for satisfying the demand directly by $v_{\rm a}(t)$ and the acquired energy to charge the energy storage by $v_{\rm b}(t)$. $M$ and $m$ can be estimated in advance, for example, based on historical data. Note that the proposed algorithm still applies, even when $M$ and $m$ are not known in advance.

\item
\textbf{Energy Storage:} The energy storage can reduce the total electricity cost by exploiting price fluctuations and harnessing excessive renewable energy. The energy storage has a capacity $B$. The level of energy storage at time slot $t$ is given by:
\begin{equation}
x(t+1)=x(t)+\eta_{\rm c}\big(r_{\rm b}(t)+v_{\rm b}(t)\big)  -\eta_{\rm d}d(t)
\end{equation}
where $d(t)$ is the energy discharged from the energy storage, $v_{\rm b}(t)$ and $r_{\rm b}(t)$ are the energy charged to the energy storage from the grid and renewable energy, respectively. $\eta_{\rm c}\le1$ and $\eta_{\rm d}\ge1$ are charging and discharging efficiencies, respectively. To capture the limitations in the charging and discharging rates, it is required that $ d(t)\le \mu_{\rm d}$ and $r_{\rm b}(t)+v_{\rm b}(t)\le \mu_{\rm c}$. The level of energy storage is required to satisfy the boundary conditions, $x(0)=B_{\rm s}$ and $x(T)=B_{\rm e}$.

\item
\textbf{Excessive Renewable Energy:} Consider arbitrary excessive renewable energy (after being offset by the original demand at the current time slot), denoted by $r(t)$. The energy to charge the energy storage is denoted by $r_{\rm b}(t)\ ( \le r(t))$. The cost of using renewable energy is assumed to be free, and hence, is most preferred to be consumed for satisfying demand. Let the ratio of total excessive renewable energy vs total excessive demand, normalized by charging and discharging efficiencies be $\rho \triangleq \frac{\eta_{\rm c}}{\eta_{\rm d}} \cdot \frac{\sum_{t=1}^T r(t)}{\sum_{t=1}^T a(t)}$, which is an effective measure of the availability of excessive renewable energy that can be stored for satisfying future demands. $\rho$ can be estimated in advance, because the long-term average is more predictable. Note that the proposed algorithm still applies, even when $\rho$ is not known in advance.

\end{itemize}

\subsection{Problem Definition}

The energy storage problem ({\sf ESP}) is formulated as follows: 
\begin{eqnarray}
({\sf ESP}) \  \min &  & \sum_{t=1}^{T}p(t)\big(v_{\rm a}(t)+v_{\rm b}(t)\big)\\
\textrm{s.t.} &  & \quad x(t+1)-x(t) \notag \\
&& =\eta_{\rm c}\big(r_{\rm b}(t)+v_{\rm b}(t)\big)  -\eta_{\rm d}d(t)\\
 &  & d(t)+v_{\rm a}(t) = a(t) \label{eqn:demand-balance}\\
 &  & r_{\rm b}(t)\le r(t) \label{eqn:renewable-balance}\\
 &  & 0\leq x(t)\leq B \label{eqn:over-underflow}\\
 &  & r_{\rm b}(t)+v_{\rm b}(t)\le \mu_{\rm c} \label{eqn:charge-rate}\\
 &  & d(t)\le \mu_{\rm d} \label{eqn:discharge-rate}\\
 &  & x(0)=B_{\rm s},\ x(T)=B_{\rm e} \label{eqn:boundary-cond}\\
 \textrm{var.} &  & x(t) \ge 0, d(t) \ge 0, r_{\rm b}(t) \ge 0, v_{\rm a}(t) \ge 0, v_{\rm b}(t) \ge 0 \notag
\end{eqnarray}

Energy storage systems may also bear other tear-and-wear and long-term maintenance costs. As in the extant literature \cite{RuiZhang15, KHT11, Gamal11}, this paper considers the short-term operation of energy storage systems, such that the electricity cost considerably outweighs the maintenance costs. Although this paper does not explicitly consider local power generators, local power generators with linear generation cost can be easily modelled as a part of the market price of the grid.

Let the inputs of the problem (i.e., the sequence of demand, market prices and renewable energy supplies) be $\sigma= \big( a(t), p(t), r(t)\big)_{t=1}^T$. 
The problem $({\sf ESP})$ can be solved by linear programming, when all inputs $\sigma$ are given in advance.

However, $\sigma$ is revealed gradually over time in reality, which requires decisions to be made without future information.
An algorithm is called {\em online}, if the decision at the current time only depends on the instantaneous information before or at the current time slot $t_{\rm now}$, namely, $\big( a(t), p(t), r(t) \big)_{t\le t_{\rm now}}$. 
Given input $\sigma$, let ${\sf Cost}({\cal A}[\sigma])$ be the cost of algorithm ${\cal A}$, and ${\sf Opt}(\sigma)$ be the cost of an offline optimal solution (that may rely on an oracle to obtain all future inputs). In competitive analysis for online algorithms \cite{Borodin:1998}, {\em competitive ratio} is a common performance metric, defined as the {\em worst-case} ratio between the cost of the online algorithm ${\cal A}$ and that of an offline optimal solution, namely,
\begin{equation}
{\sf CR}({\cal A}) \triangleq \max_{\sigma} \frac{{\sf Cost}({\cal A}[\sigma])}{{\sf Opt}(\sigma)}
\end{equation}
This paper provides competitive online algorithms for solving ${\sf ESP}$ with a worst-case guarantee.

\section{Online Algorithm}

Algorithm ${\cal A}_{\sf thb}$ is a simple threshold based online algorithm, which stores energy from the grid at each time slot $t$ up to level $\hat{B}$ when the market price is below the threshold $\theta$. Otherwise, it discharges from the energy storage to satisfy the demand, if available. Meanwhile, it stores any excessive renewable energy, subject to the charging rate constraint. Let $\lbrack\ \cdot\ \rbrack^+$ be the positivity operator ($\max\{\ \cdot\ , 0 \}$).

\begin{algorithm}[htb!]
\caption{${\cal A}_{\sf thb}[\theta, \hat{B}, t, a(t), p(t), r(t)]$}
\begin{algorithmic}[1]
\LeftComment{{\em Store energy from renewable energy}}
\State $r_{\rm b}(t) \leftarrow \min\big\{r(t),\  \frac{B-x(t-1)}{\eta_{\rm c}},\  \mu_{\rm c} \big\}$
\If{$p(t) \le \theta$}
\LeftComment{{\em No discharging from energy storage}}
\State $d(t) \leftarrow 0$
\LeftComment{{\em Satisfy demand from grid}}
\State $v_{\rm a}(t) \leftarrow a(t)$ 
\LeftComment{{\em Store energy up to $\hat{B}$ from grid}}
\State $v_{\rm b}(t) \leftarrow \min\big\{ [\frac{\hat{B}-x(t-1)}{\eta_{\rm c}}-r_{\rm b}(t)]^+, [\mu_{\rm c}-r_{\rm b}(t)]^+\big\}$
\Else
\LeftComment{\mbox{\em Discharge from energy storage}}
\State $d(t) \leftarrow \min\{a(t),\ \mu_{\rm d},\ \frac{x(t-1)}{\eta_{\rm d}}\}$
\LeftComment{{\em Satisfy remaining demand from grid}}
\State $v_{\rm a}(t) \leftarrow a(t)-d(t)$
\LeftComment{{\em Store no energy from grid}}
\State $v_{\rm b}(t) \leftarrow 0$
\EndIf
\LeftComment{{\em Obtain energy storage level}}
\State $x(t) \leftarrow x(t-1) + \eta_{\rm c}\big(r_{\rm b}(t)+v_{\rm b}(t)\big)  -\eta_{\rm d}d(t)$
\State \Return $\big(x(t), d(t), r_{\rm b}(t), v_{\rm a}(t), v_{\rm b}(t)\big)$
\end{algorithmic}
\end{algorithm}

It is critical to set the parameters ($\theta, \hat{B}$) of ${\cal A}_{\sf thb}$ properly. A small $\theta$ or $\hat{B}$ will make an aggressive algorithm favoring low market prices and renewable energy, but may end up paying more electricity cost if there is insufficient energy in the storage to satisfy the demand. On the contrary, a large $\theta$ or $\hat{B}$ will make a conservative algorithm that may miss low market prices and renewable energy, because the energy storage may be saturated too soon. A proper setting of $\theta$ and $\hat{B}$ should balance the two extreme cases. Theorem~\ref{thm:online-comp3} presents a plausible setting of $\theta$ and $\hat{B}$, supported by a competitive ratio as the worst-case guarantee.

\medskip

\begin{customthm}{1} \label{thm:online-comp3}
Suppose the terminal condition $x(T) = B$ and $\rho \le 1$.
Setting $\hat{B} = B(1-\rho)$ and
\begin{equation}
\theta = \frac{\sqrt{\rho^2 (M - m)^2 + 4 M m} - \rho ( M - m) }{2} \cdot \frac{\eta_{\rm c}}{\eta_{\rm d}},
\end{equation}
where $\varphi=\frac{M}{m}$, the competitive ratio of algorithm ${\cal A}_{\sf thb}$ is 
\begin{equation}
{\sf CR}({\cal A}_{\sf thb}) = \frac{1}{2} (\rho \varphi + \rho + \sqrt{ 4 \varphi + \rho^2 (\varphi - 1)^2})
\end{equation}
\end{customthm}

\medskip

The proof can be found in the Appendix. The basic idea is that we first decompose the demand $a(t)$ into a set of simple demands called ``one-shot demand''. For each one-shot demand, we characterize the outcome by two opposite cases: (1) charging the energy storage from the grid according at the threshold, (2) not charging the energy storage from the grid because of higher-than-threshold market prices. An adversary will select the worst among the two cases. In order to minimize the competitive ratio given $\hat{B} = B(1-\rho)$, we select the threshold $\theta$, such that the optimality ratios of the two cases are equivalent, by solving a quadratic equation.

{\bf Remarks}: There are several remarks to Theorem~\ref{thm:online-comp3}.
\begin{enumerate}

\item
When $\rho = 0$, then $\hat{B} = B$ and $\theta = \sqrt{M m}\cdot\frac{\eta_{\rm c}}{\eta_{\rm d}}$. The competitive ratio becomes ${\sf CR}({\cal A}_{\sf thb}) = \sqrt{\varphi}$. 

When $\rho = 1$, then $\hat{B} = 0$ and $\theta = m\cdot\frac{\eta_{\rm c}}{\eta_{\rm d}}$, which is equivalent to never storing energy from the grid. The competitive ratio becomes ${\sf CR}({\cal A}_{\sf thb}) = \varphi + 1$. 

One can show that $m\cdot\frac{\eta_{\rm c}}{\eta_{\rm d}} \le \theta \le \sqrt{M m}\cdot\frac{\eta_{\rm c}}{\eta_{\rm d}}$ and $\sqrt{\varphi} \le {\sf CR}({\cal A}_{\sf thb}) \le \varphi + 1$.
The threshold setting in Theorem~\ref{thm:online-comp3} offers a continuous transition from $\sqrt{\varphi}$ (when $\rho = 0$) to $\varphi + 1$ (when $\rho = 1$). Note that the competitive ratio varies almost linearly as $\rho$, namely ${\sf CR}({\cal A}_{\sf thb}) = \Theta(\rho)$ (see Fig.~\ref{fig:funcs} for an illustration). 

\begin{figure}[htb!]
\hspace{-20pt}
\includegraphics[scale=0.8]{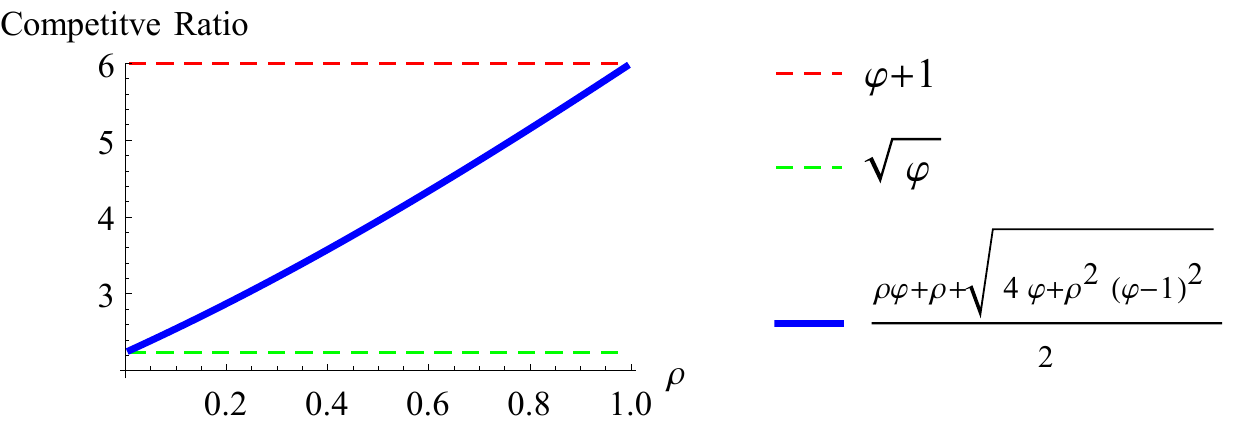} 
\caption{An illustration of competitive ratio with $\varphi = 5, \frac{\eta_{\rm c}}{\eta_{\rm d}} = 1$.} \label{fig:funcs} 
\end{figure}

\item Competitive ratio ${\sf CR}({\cal A}_{\sf thb})$ characterizes the worst-case guarantee of optimality. In practice, the observed optimality ratio of ${\cal A}_{\sf thb}$ can be far better than the worst-case upper bound. It is observed from simulations in Sec.~\ref{sec:sim} that the empirical optimality ratio of ${\cal A}_{\sf thb}$ is well below the competitive ratio.

\item When $\rho > 1$, we apply Theorem~\ref{thm:online-comp3} as if $\rho = 1$. Namely, we set $\hat{B} = 0$ and $\theta = m\cdot\frac{\eta_{\rm c}}{\eta_{\rm d}}$. The competitive ratio is still ${\sf CR}({\cal A}_{\sf thb}) = \varphi + 1$. 

\item There is an equivalence between the terminal condition $ x(T) = B_{\rm e}$ and demand $a(T)$ at the final timeslot. One 
can substitute terminal condition $x(T) = B_{\rm e}$ by an additional demand $a(T) = \eta_{\rm c} B_{\rm e}$, or vice versa. When the terminal condition becomes $ x(T) = B_{\rm e} < B$, one can still apply Theorem~\ref{thm:online-comp3} by substituting $\rho$ by $\tilde{\rho} = \frac{\eta_{\rm c}}{\eta_{\rm d}} \cdot \frac{B- B_{\rm e} + \sum_{t=1}^T r(t)}{\sum_{t=1}^T a(t)}$ in Theorem~\ref{thm:online-comp3}. The basic idea is that at most $\eta_{\rm c}(B - B_{\rm e})$ of the demand may be eliminated without incurring any cost, which is similar to the effect that the demand is satisfied by renewable energy. However, it is observed from simulations in Sec.~\ref{sec:sim} that the boundary conditions have a marginal effect on the empirical optimality ratio of ${\cal A}_{\sf thb}$, if $T$ is relatively large.

\item In case that the values of $M$, $m$, $\rho$ are not known in advance, ${\cal A}_{\sf thb}$ can be adapted to estimate these parameters dynamically. First, set $M = m = p(1)$. Then $M$ and $m$ are updated to be the maximum and minimum market prices observed so far at time slot $t$. $\rho$ can also be estimated by the observed normalized ratio of renewable energy over total demand at time slot $t$. If $T$ is relatively large, the estimated $M$, $m$, $\rho$ will converge to the true values. The empirical performance is studied by simulations in Sec.~\ref{sec:sim}.

\end{enumerate}

The following theorems provide the lower bounds of competitive ratio of any deterministic online algorithm for the extreme cases $\rho=0$ and $\rho=1$. 
Hence, threshold based algorithm ${\cal A}_{\sf thb}$ attains a competitive ratio that is within a constant factor of the lower bounds.

\medskip

\begin{customthm}{2}  \label{thm:lb-det0}
Consider zero excessive renewable energy ($\rho=0$). The competitive ratio of any deterministic online algorithm ${\cal A}$ is ${\sf CR}({\cal A}) \ge \frac{1}{2}\big(1+\sqrt{\varphi}\big)$.
\end{customthm}

\medskip

\begin{customthm}{3} \label{thm:lb-det1}
Consider abundant excessive renewable energy ($\rho\ge1$). The competitive ratio of any deterministic online algorithm ${\cal A}$ is ${\sf CR}({\cal A}) \ge \varphi$.
\end{customthm}
\section{Lookahead Algorithm}

In certain circumstances, limited future information for a lookahead window of size $W$ (i.e., $\big(a(\tau), p(\tau), r(\tau)\big)_{\tau=t}^{t+W})$) may be available, possibly due to predictions of the near future market prices, renewable energy and demands. While this paper does not investigate the prediction mechanisms, we can adapt our threshold based online algorithm to take advantage of the limited future information in a sliding window manner.

\begin{algorithm}[htb!]
\caption{${\cal A}_{\sf thb}^{\sf lka}\Big[\theta, \hat{B}, t, x(t-1), \big(a(\tau), p(\tau), r(\tau)\big)_{\tau=t}^{t+W}\Big]$}
\begin{algorithmic}[1]
\LeftComment{{\em Store renewable energy as much as possible}}
\For{$\tau \in [t, t+W]$}
\State $r_{\rm b}(\tau) \leftarrow \min\big\{r(\tau),\  \frac{B-x(\tau-1)}{\eta_{\rm c}},\  \mu_{\rm c} \big\}$
\State $x(\tau) \leftarrow x(\tau-1) + \eta_{\rm c} r_{\rm b}(\tau)$
\EndFor
\LeftComment{{\em Obtain offline optimal solution within window $W$}}
\State $\big(x(\tau), d(\tau), v_{\rm a}(\tau), v_{\rm b}(\tau)\big)_{\tau=t}^{t+W}$ 
\Statex $\quad \leftarrow {\cal A}_{\sf ofl}\Big[x(t-1), \big(a(\tau), p(\tau), x(\tau)\big)_{\tau=t}^{t+W}\Big]$
\LeftComment{{\em Find min-price time slot within window $W$}}
\If{$t = \arg\min_{ t \le \tau \le t+W}\big\{ p(\tau)\big\}$ and $p(t) \le \theta$}
\LeftComment{{\em Find residual capacity w.r.t. offline decisions}}
\State $\tilde{x} \leftarrow \max_{t \le \tau \le t+W}\big\{ x(\tau)\big\}$
\State ${y} \leftarrow \min\big\{ B - \tilde{x}, [\hat{B} - x(t+W)]^+\big\}$
\LeftComment{{\em Store energy from grid}}
\State $v_{\rm b}(t) \leftarrow \min\big\{ [\frac{{y}}{\eta_{\rm c}} - r_{\rm b}(t) + d(t)]^+, [\mu_{\rm c} - r_{\rm b}(t)]^+\big\}$
\State $x(t) \leftarrow x(t-1) + \eta_{\rm c}\big(r_{\rm b}(t)+v_{\rm b}(t)\big)  -\eta_{\rm d}d(t)$
\EndIf
\State \Return $\big(x(t), d(t), r_{\rm b}(t), v_{\rm a}(t), v_{\rm b}(t)\big)$
\end{algorithmic}
\end{algorithm}

We present lookahead threshold based algorithm ${\cal A}_{\sf thb}^{\sf lka}$ as a heuristic to integrate the offline optimal solution within lookahead window $W$ and the threshold based online decisions of charging the energy storage. The computations are carried out in a sliding window fashion. Namely, at every time slot $t$, the decisions are computed from $t$ to $t+W$, but only the decision at $t$ is applied.

Let ${\cal A}_{\sf ofl}$ be the algorithm for computing offline optimal solution within a lookahead window $W$, given the initial energy storage level $x(t-1)$ and inputs $\big(a(\tau), p(\tau), x(\tau)\big)_{\tau=t}^{t+W}$. ${\cal A}_{\sf ofl}$ assumes that the renewable energy has been properly stored in the energy storage beforehand, and hence does not consider renewable energy.

The decisions are computed as follows. First, it stores the renewable energy as much as possible within lookahead window $W$. Second, it invokes ${\cal A}_{\sf ofl}$ to obtain the optimal solution within lookahead window $W$. Third, it stores energy up to the level $\hat{B}$ from the grid at the current timeslot $t$, if the market price at $t$ is the minimum within the next $W$ timeslots and is below the threshold $\theta$. To avoid the interference with the offline decisions, the amount of stored energy at $t$ is less than $B- \tilde{x}$ and $[\hat{B}-x(t+W)]^+$, where $\tilde{x}$ is maximum level of energy storage within $[t, t+W]$ that is computed from offline optimal solution. 

The parameters ($\theta, \hat{B}$) of ${\cal A}_{\sf thb}^{\sf lka}$ are set according to Theorem~\ref{thm:online-comp3}.
It is observed from simulations in Sec.~\ref{sec:sim} that the empirical ratio between ${\cal A}_{\sf thb}^{\sf lka}$ and the offline optimal solution can be significantly improved when $W$ increases.

\section{Empirical Evaluation}\label{sec:sim}

\begin{figure*}[htb!] \vspace{-10pt}
    \begin{minipage}[b]{0.5\textwidth}
	\centering
            \includegraphics[scale = 0.22]{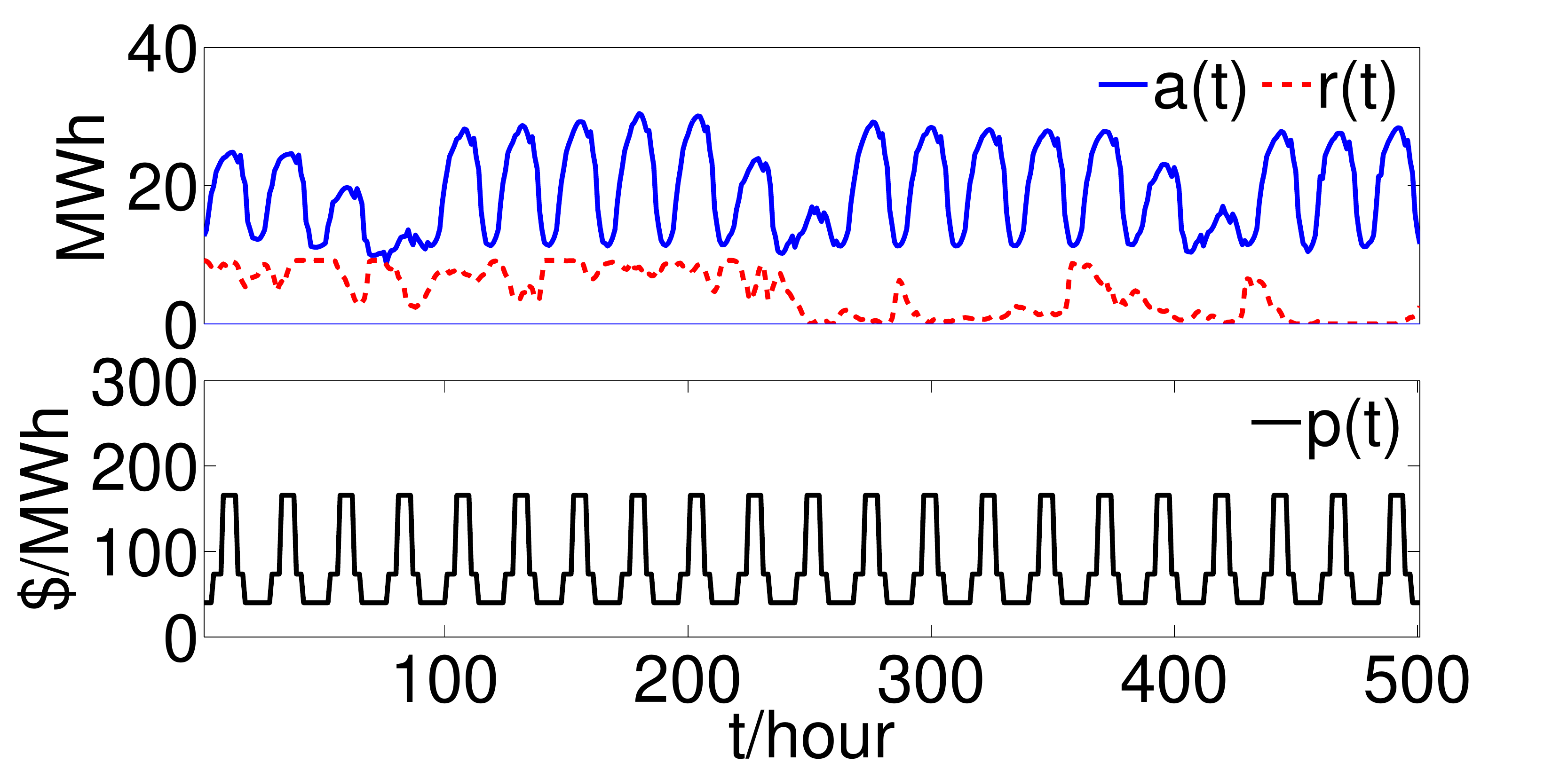}            
        \caption{Data traces of demand, renewable energy, market prices.}\label{fig:data} 
    \end{minipage}	
    \begin{minipage}[b]{0.5\textwidth} 
	\centering 
            \includegraphics[scale = 0.22]{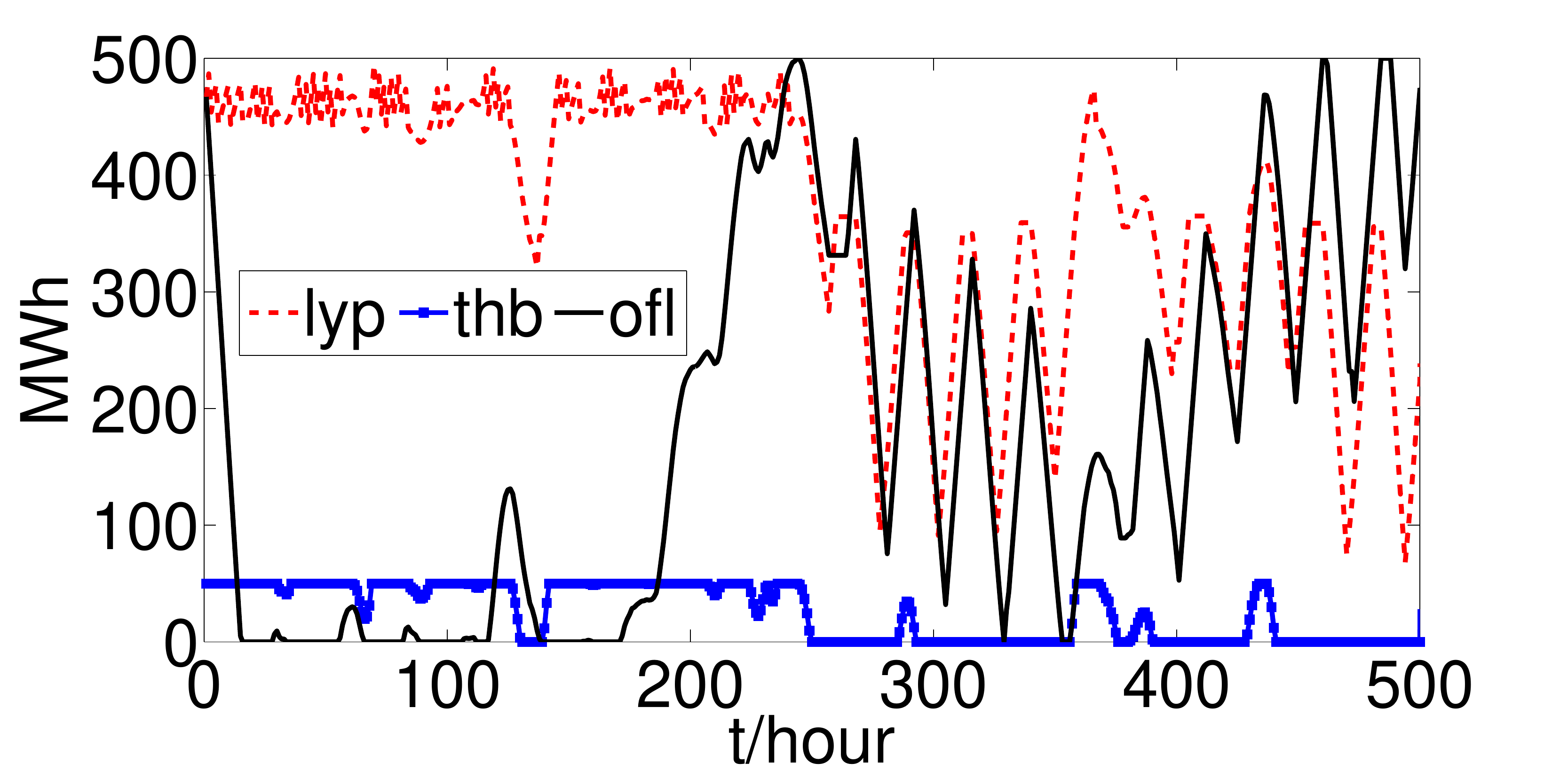}            
        \caption{Energy storage level $x(t)$.}\label{fig:battery_level_1} 
    \end{minipage}
\end{figure*}

\begin{figure*}[htb!] \vspace{-10pt}
     \begin{minipage}[b]{0.5\textwidth}
	 \centering
            \includegraphics[scale = 0.22]{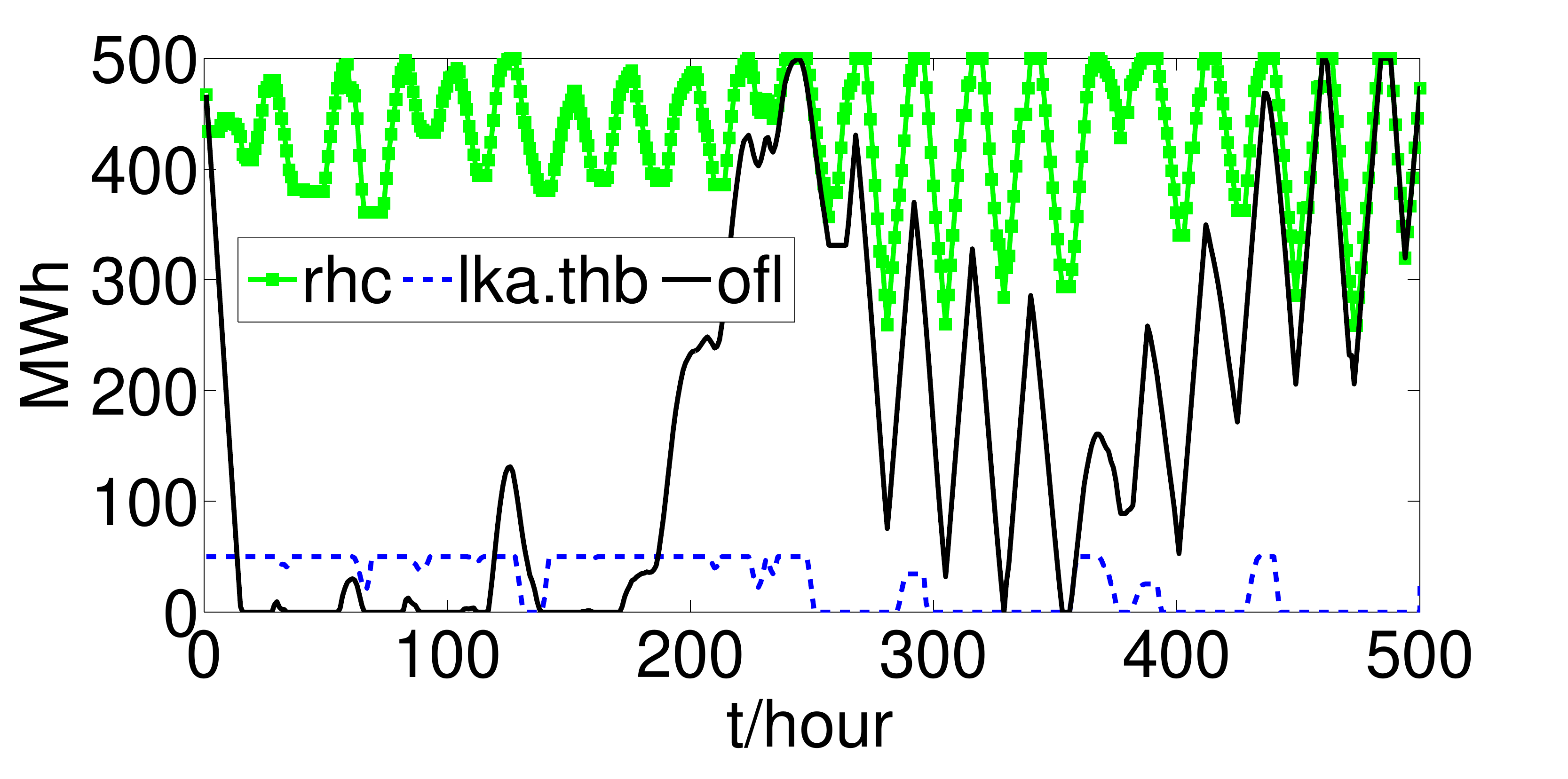}            
        \caption{Energy storage level $x(t)$.}\label{fig:battery_level_2} 
    \end{minipage}
	    \begin{minipage}[b]{0.5\textwidth}
		\centering
 \includegraphics[scale = 0.22]{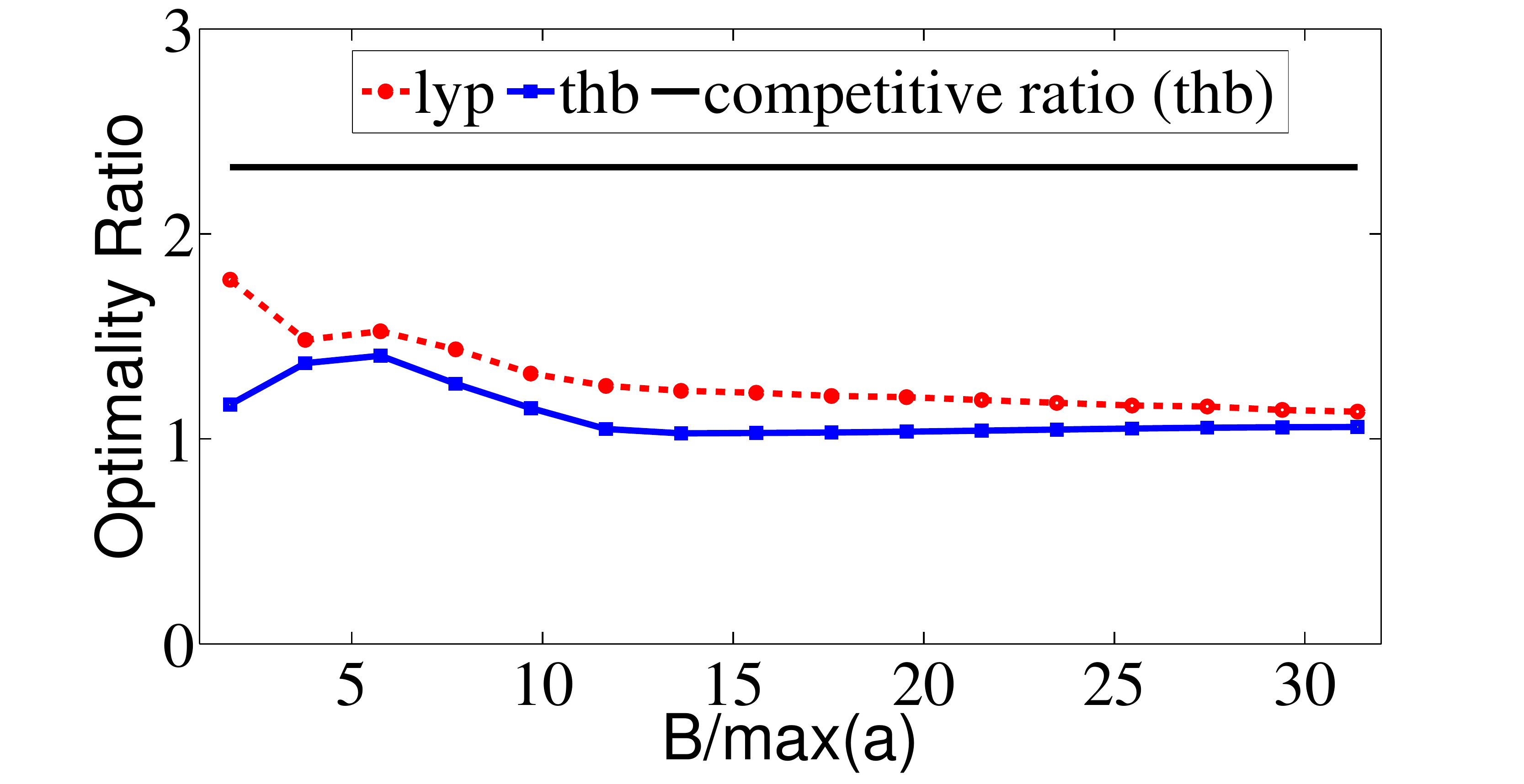}
        \caption{Empirical optimality ratio vs. competitive ratio.}\label{fig:ratio} 
    \end{minipage}
\end{figure*}

\begin{figure*}[htb!] \vspace{-10pt}
	     \begin{minipage}[b]{0.5\textwidth}
          \centering
            \includegraphics[scale = 0.22]{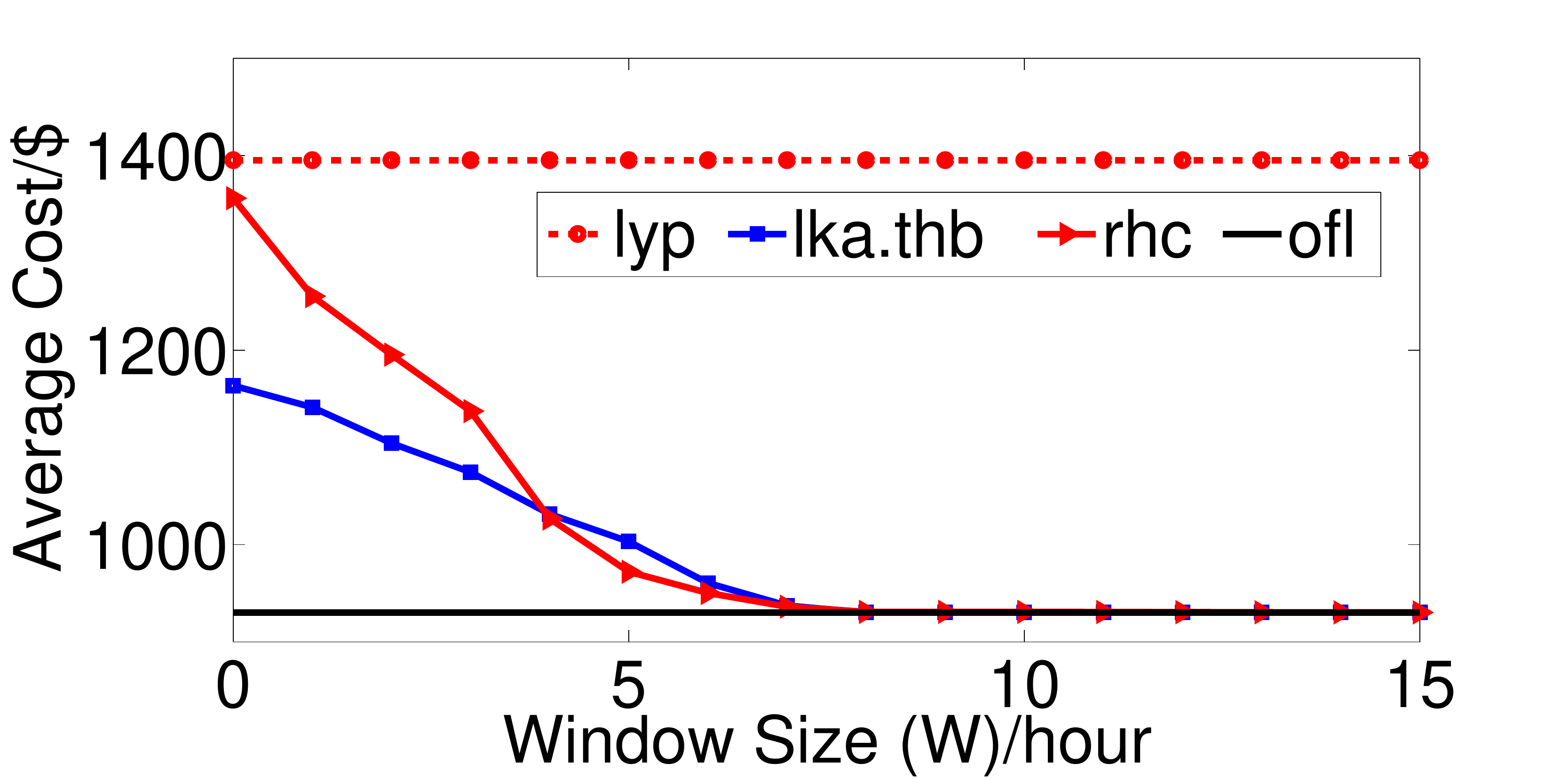}
        \caption{Average cost vs. time window size $W$.}\label{fig:lookahead}
    \end{minipage}
    \begin{minipage}[b]{0.5\textwidth}
       \centering
            \includegraphics[scale = 0.22]{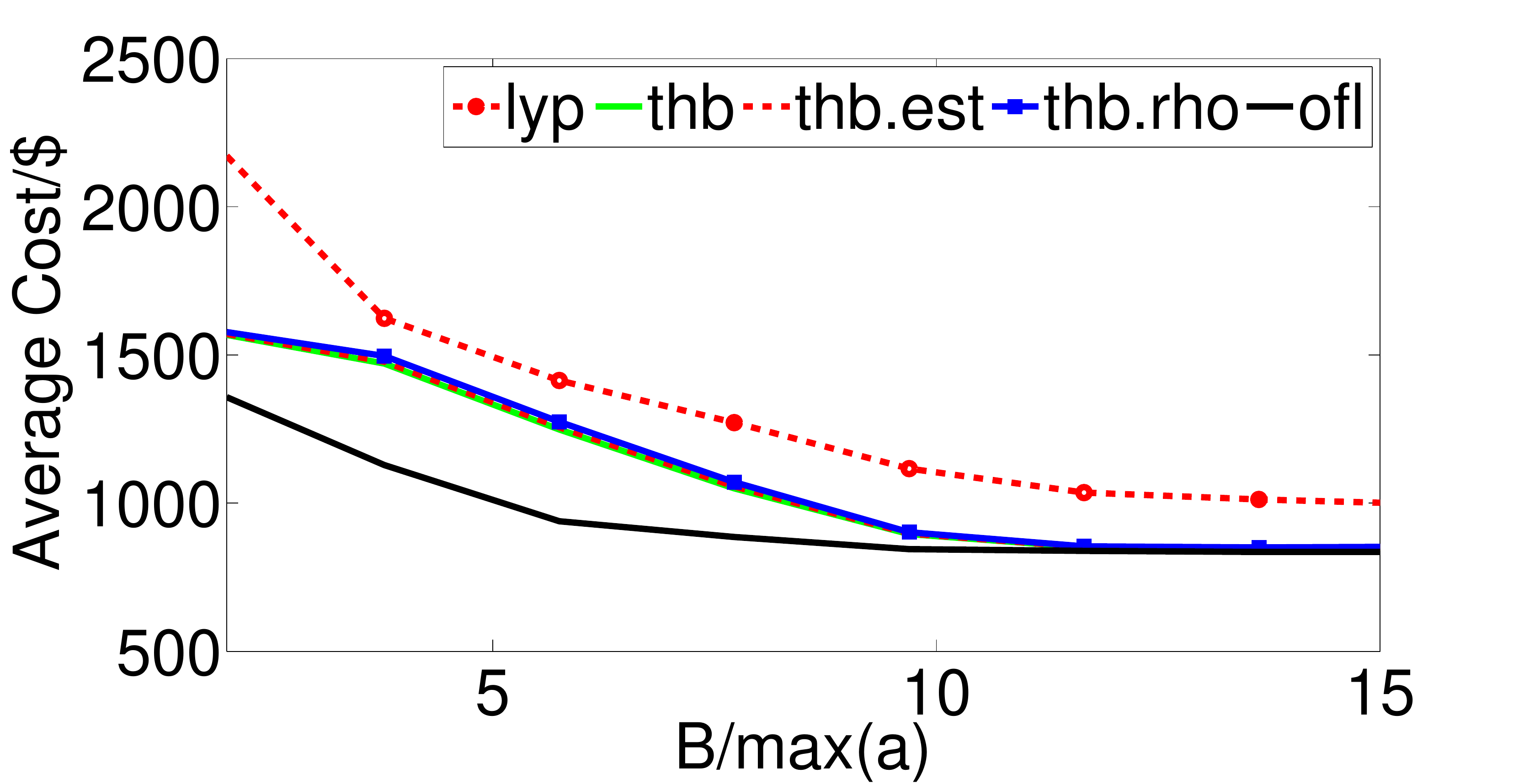}
        \caption{Average cost vs. capacity (when $x(0) = x(T) = B$).}\label{fig:cost1} 
    \end{minipage}		
\end{figure*}

The empirical optimality of the proposed algorithms is evaluated based on the simulation studies using real-world data traces. Apart form comparing with the offline optimal solution, the online algorithm is also compared with Lyapunov stochastic optimization algorithm, whereas the lookahead algorithm is compared with receding horizon control algorithm.

\subsection{Data Traces and Parameters} \label{sec:setting}

\begin{itemize}

\item
{\em Demand}: The demand traces are obtained from California Commercial End-Use Survey (CEUS) \cite{CEUS} of a college in San Francisco, which consumed about 154 GWh electricity per year. The data traces contain hourly electricity demands in year 2002. As shown in Fig.~\ref{fig:data}, regular daily pattern is observed in peak and off-peak hours, with typical weekday and weekend variations.

\item 
{\em Renewable Energy}: The wind power traces are obtained from \cite{NREL}. The power output data is used within the same time period with a resolution of 1 hour of an offshore wind farm right outside San Francisco with an installed capacity of 12MW. The wind power trace is shown in Fig.~\ref{fig:data}. Roughly speaking, the renewable energy supplies about 10\% of the total energy demand.

\item
{\em Electricity Prices}: The electricity price data are obtained from PG\&E \cite{PG_E} and are shown in Fig. \ref{fig:data}. Both the electricity demands and the prices show strong diurnal properties: in the daytime, the demands and prices are relatively high; at nights, both are low. This suggests the feasibility of reducing the operating cost by charging the energy storage from the grid during night and discharging the energy storage to satisfy demand during daytime.

\item
{\em Energy Storage}: 
The charging and discharging rate constraints $\mu_{\rm c}$ and $\mu_{\rm d}$ are set as 30MWh/h. The charging and discharging efficiencies are set set as $\eta_{\rm d}=1.1, \eta_{\rm c}=0.9$.

\end{itemize}

\subsection{Observations of Threshold Based Algorithm}

The optimality of online threshold based algorithm ${\cal A}_{\rm thb}$ (${\sf thb}$) is compared with that of offline optimal solution (${\sf ofl}$) and Lyapunov stochastic optimization algorithm (${\sf lyp}$), under different values of energy storage capacity $B$ (normalized by the peak excess demand $\max(a)$). The basic idea of Lyapunov stochastic optimization is to solve a relaxed version of {\sf ESP}, without the capacity constraint (\ref{eqn:over-underflow}). Based on the concept of Lyapunov drift, Lyapunov stochastic optimization uses perturbed parameters to limit the violation of capacity constraint, when the capacity is large, and averaged over a sufficiently long time horizon. The implementation of Lyapunov stochastic optimization algorithm follows from that in \cite{Neely11}. 
In Fig.~\ref{fig:battery_level_1}, the instantaneous energy storage level $x(t)$ for is plotted for ${\sf ofl}$, ${\sf lyp}$ and ${\cal A}_{\rm thb}$. It is observed that ${\sf lyp}$ tends to store energy from the grid at a higher level than ${\cal A}_{\rm thb}$.

In all three algorithms, the electricity cost decreases as the capacity $B$ increases, because more energy storage capacity allows more opportunities to harness renewable energy and price fluctuations.  
The observed optimality ratios of ${\cal A}_{\rm thb}$ and ${\sf lyp}$ are plotted in Fig.~\ref{fig:ratio}. The observed optimality ratio of ${\cal A}_{\sf thb}$ is far better than the worst-case upper bound in Theorem~\ref{thm:online-comp3}. It is observed that ${\cal A}_{\rm thb}$ outperforms ${\sf lyp}$, in particular with small or medium sizes of capacity $B$. When $B$ is sufficiently large, both algorithms approach the offline optimal cost. The optimality ratio of ${\cal A}_{\rm thb}$ increases initially for small $B$, because the offline optimal cost decreases more initially than the cost of ${\cal A}_{\rm thb}$ does.

\subsection{Observations of Lookahead Algorithm}

The optimality of lookahead threshold based algorithm ${\cal A}_{\rm thb}^{\sf lka}$ (${\sf lka.thb}$) is compared with that of offline optimal solution (${\sf ofl}$) and receding horizon control algorithm (${\sf rhc}$), under different values of energy storage capacity $B$ (normalized by the peak demand $\max(a)$). The receding horizon control algorithm optimizes the energy storage management decisions within a lookahead window $W$, and proceeds in a sliding window manner as time goes on. The key difference between ${\sf rhc}$ and ${\sf lka.thb}$ is that ${\sf lka.thb}$ makes online decisions to acquire extra energy from the grid to charge the energy storage (using the same market price threshold as ${\cal A}_{\rm thb}$), whereas ${\sf rhc}$ does not consider any online decisions.

Both ${\cal A}_{\rm thb}^{\sf lka}$ and ${\sf rhc}$ are compared with the same time window size $W$. In Fig.~\ref{fig:battery_level_2}, the instantaneous energy storage level $x(t)$ for is plotted for ${\sf ofl}$, ${\sf rhc}$ and ${\cal A}_{\rm thb}^{\sf lka}$. The cost averaged by $T$ of ${\cal A}_{\rm thb}^{\sf lka}$, offline optimal solution and ${\sf rhc}$ are plotted in Fig.~\ref{fig:lookahead}. ${\cal A}_{\rm thb}^{\sf lka}$ is a heuristic that integrates offline computation within lookahead window $W$ (like ${\sf rhc}$) and threshold based online decisions to charge energy storage (like ${\cal A}_{\rm thb}$). It is observed that the cost achieved by ${\cal A}_{\rm thb}^{\sf lka}$ is significantly lower than ${\sf rhc}$, when $W$ is small. ${\sf rhc}$ only outperforms ${\cal A}_{\rm thb}^{\sf lka}$ slightly, when $W$ is larger. But ${\cal A}_{\rm thb}^{\sf lka}$ approaches the offline optimal cost closely, when $W$ becomes about 8 hours.

\subsection{Impact of Unknown Parameters and Boundary Conditions}

When the parameters of $M$, $m$, or $\rho$ is not known a-priori, ${\cal A}_{\sf thb}$ is adapted to estimate these parameters dynamically. In Fig.~\ref{fig:cost1}, ${\sf thb.est}$ estimates $M$ and $m$ by the maximum and minimum market prices observed so far at time $t$, whereas ${\sf thb.rho}$ estimates $\rho$ by the observed normalized ratio of renewable energy over total demand at time $t$. It is observed that the estimated $M$, $m$, $\rho$ converge fast to the true values. Both ${\sf thb.est}$ and ${\sf thb.rho}$ behave similarly as ${\cal A}_{\rm thb}$ (that knows $M$, $m$ and $\rho$ a-priori).

\begin{figure}[!htb]
\centering \vspace{-10pt}
 \includegraphics[scale = 0.22]{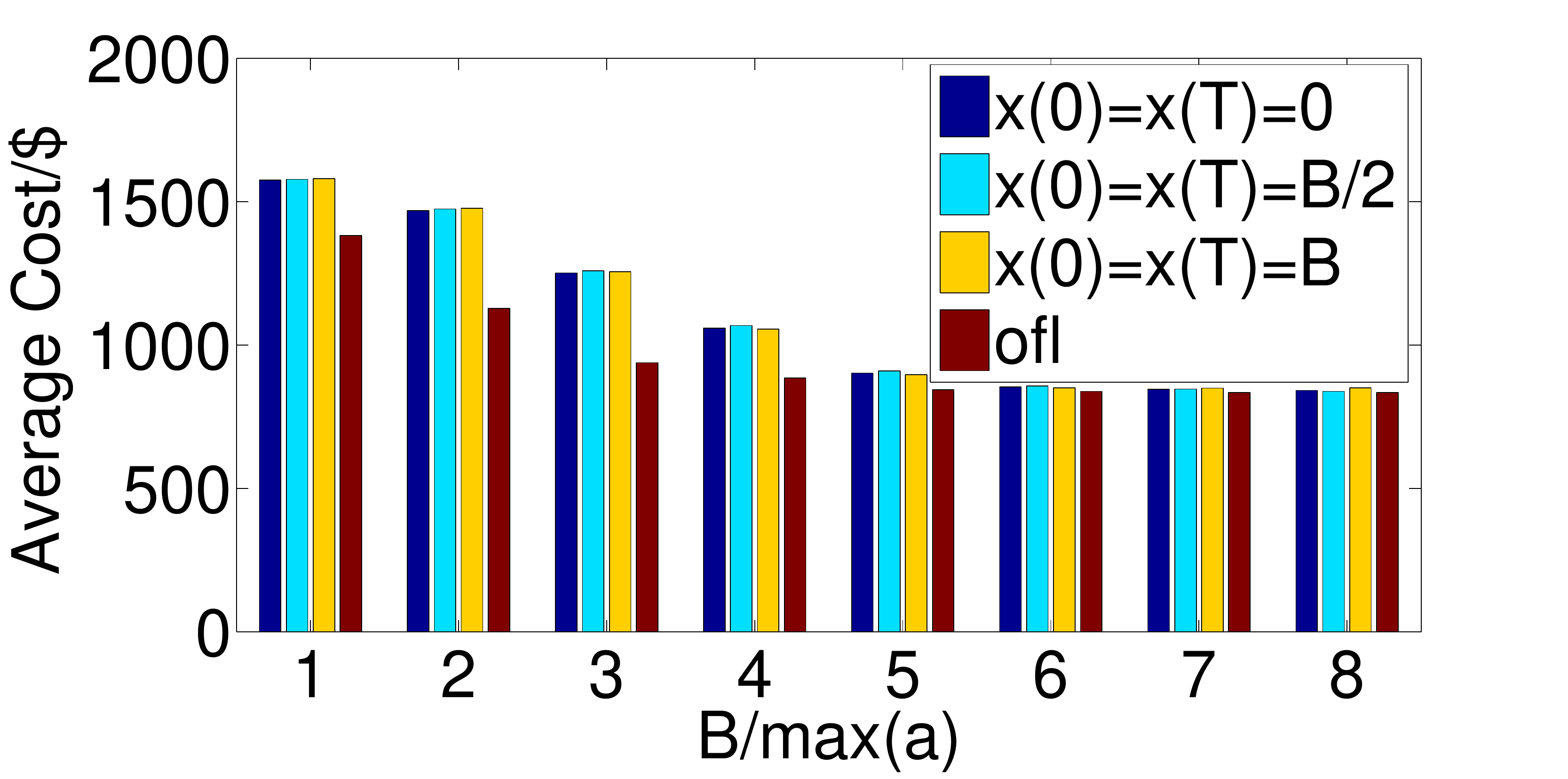}
\caption{\label{fig:cost2}The average cost of ${\sf thb.est}$ under different boundary conditions, as compared with {\sf ofl} (when $x(0)=x(T)=0$).}
\end{figure}

\begin{figure}[!htb]
\centering \vspace{-20pt}
 \includegraphics[scale = 0.22]{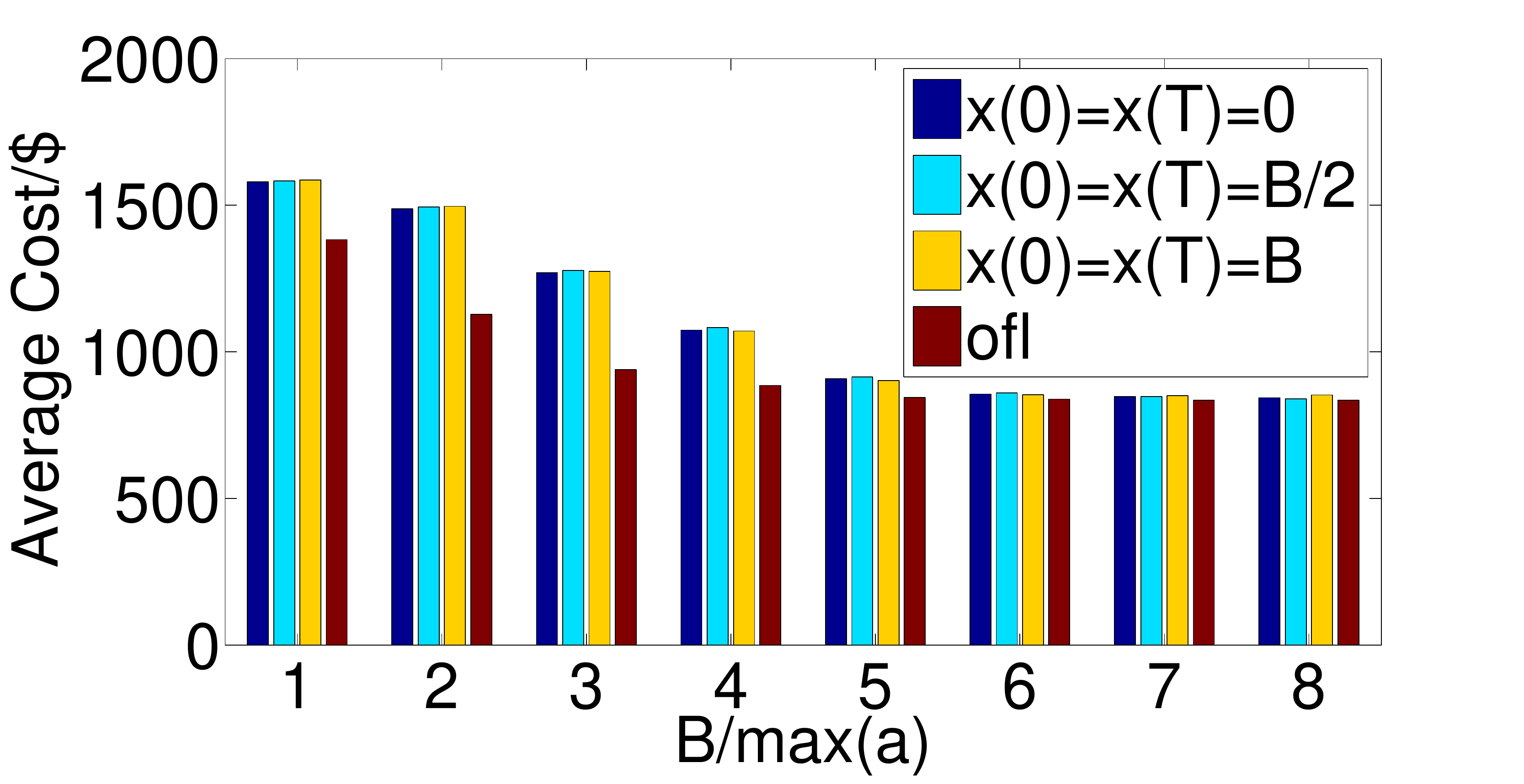}
\caption{\label{fig:cost3}The average cost of ${\sf thb.rho}$ under different boundary conditions, as compared with {\sf ofl} (when $x(0)=x(T)=0$).}
\end{figure}

The impacts of boundary conditions $x(0)=B_{\rm s}$ and $x(T)=B_{\rm e}$ are studied in Figs.~\ref{fig:cost2}-\ref{fig:cost3}, with three different boundary conditions are considered: (1) $x(0) = x(T) = B$, (2) $x(0) = x(T) = 0$, (3) $x(0) = x(T) = \frac{B}{2}$. It is observed that the boundary conditions have a marginal effect on the empirical optimality ratio of ${\cal A}_{\sf thb}$, because when $T$ is relatively large, the demand has the dominant effect.

\subsection{Impact of Charging and Discharging Rate Constraints}

\begin{figure}[!htb]
\centering \vspace{-10pt}
\includegraphics[scale = 0.22]{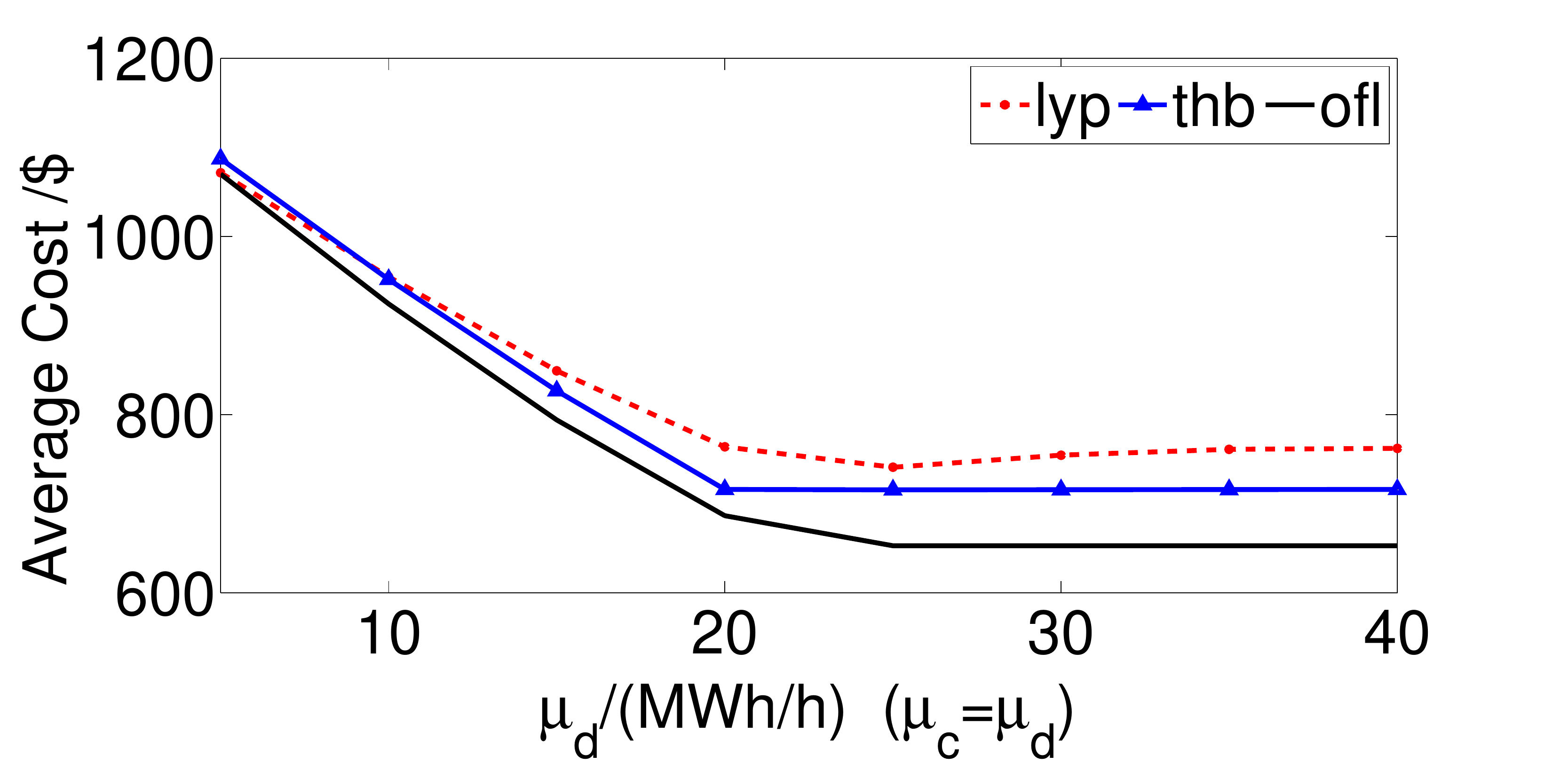} \vspace{-10pt}
\caption{Average cost vs. $\mu_{\rm d}$ and $\mu_{\rm c}$.} \label{fig:cdis}
\end{figure}

The impact of charging and discharging rate constraints ($\mu_{\rm c}$ and $\mu_{\rm d}$) is studied in Fig.~\ref{fig:cdis}, where $\mu_{\rm d}$ is varied from 5MWh/h to 40MWh/h and $\mu_{\rm c}=\mu_{\rm d}$. All three algorithms perform better with increasing $\mu_{\rm d}$ and $\mu_{\rm c}$. When $\mu_{\rm d}$ increases to about 20MWh/h, the costs achieved by the algorithms reach the minimum and do not decrease any more. This is because 20MWh is roughly the peak power demand of this trace. The maximum benefit of energy storage is attained by discharging to satisfy the peak power demand when the market price attains the maximum. Consequently, with a maximum discharging rate of larger than the peak demand, the cost cannot be further reduced. 

\subsection{Impact of Charging and Discharging Efficiencies}

\begin{figure}[!htb]
\centering \vspace{-10pt}
\includegraphics[scale = 0.22]{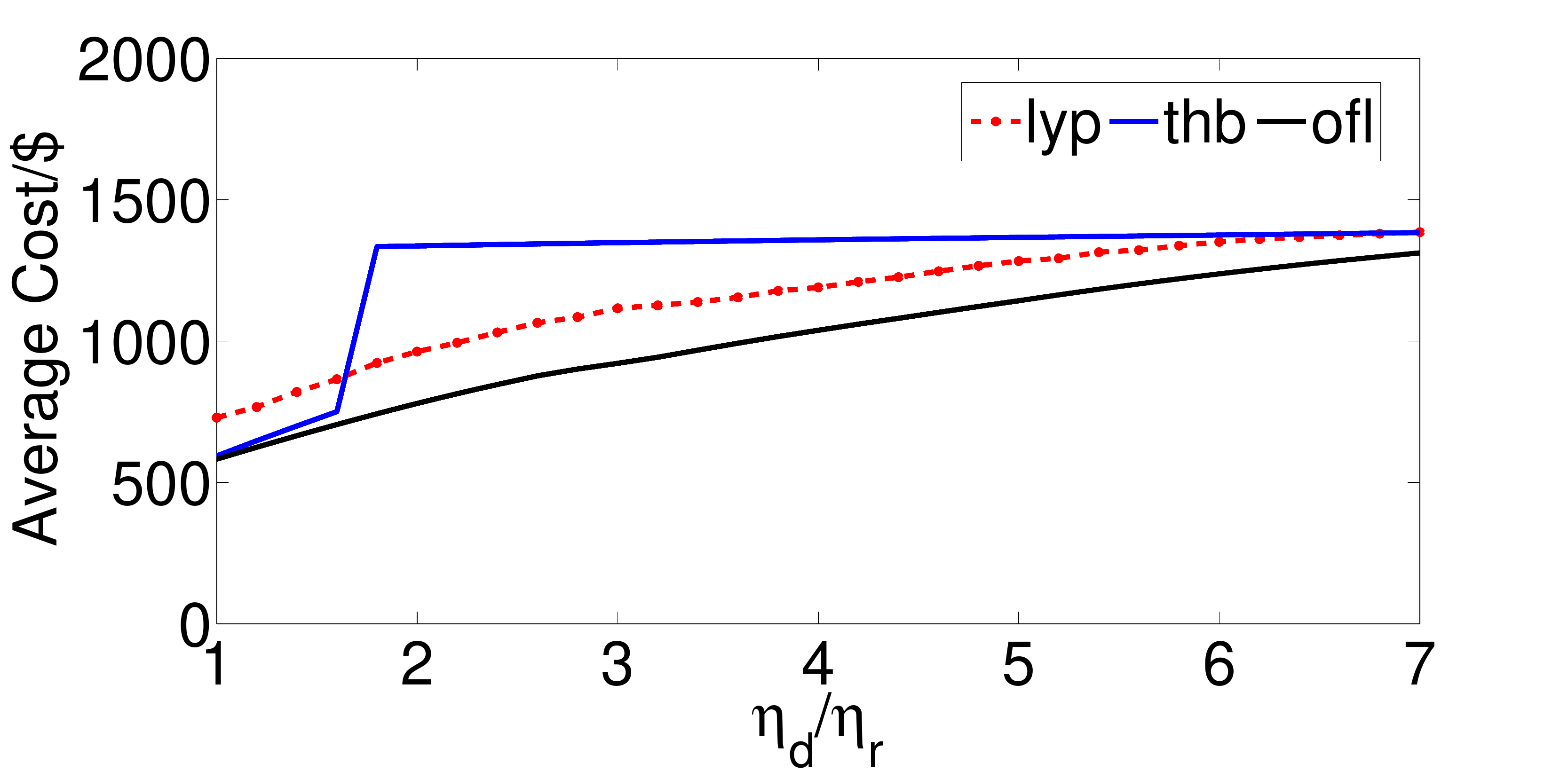} \vspace{-10pt}
\caption{Average cost vs. $\frac{\eta_{\rm d}}{\eta_{\rm c}}$.} \label{fig:eff}
\end{figure}

The impact of charging and discharging efficiencies ($\eta_{\rm c}$ and $\eta_{\rm d}$) is studied in Fig.~\ref{fig:eff}, where the ratio $\frac{\eta_{\rm d}}{\eta_{\rm c}}$ is varied from 1 to 7. The costs obtained by the three algorithms increase monotonically with increasing $\frac{\eta_{\rm d}}{\eta_{\rm c}}$. It is observed that ${\cal A}_{\rm thb}$ performs close to the offline optimal solution when $\frac{\eta_{\rm d}}{\eta_{\rm c}} \le 1.5$, which is a reasonable setting in most energy storage.

\section{Conclusion} \label{sec:concl} 

This paper studied the online algorithms for energy storage management in the presence of price fluctuations and renewable energy sources, for minimizing the electricity cost from the grid. Competitive online algorithms are devised that can cope with arbitrary inputs, any energy storage capacity and finite time horizon. It is observed that the proposed algorithms can outperform the other extant algorithms, such as Lyapunov optimization and receding horizon control algorithms. In contrast to the deterministic online algorithms studied in this paper, one can also consider randomized online algorithms for energy storage management. However, it is shown that randomized online algorithms cannot outperform deterministic online algorithms for one-way minimum trading problem in terms of the order magnitude of competitive ratio \cite{lorenz2009optimal}. Therefore, randomized online algorithms cannot give a superior performance in the more general energy storage management problem considered in this paper. In future work, the presence of local power generators with non-linear generation cost will be considered.

\bibliographystyle{ieeetr}
\bibliography{reference}

\appendix

\subsection{1-Min Search and One-way Trading Problems}

In 1-min search problem, a trader is searching for the minimum price of some asset. At each time slot $t$, the trader is presented a price $p(t)$ and must decide whether or not to accept this price. Once the trader decides to accept the price $p(t)$, the search ends and the trader's cost is $p(t)$. If the trader does not accept any price for the first $T-1$ time slots, he needs to accept any price at time slot $T$. 
According to \cite{Yaniv01}, the online algorithm that accepts the first price below threshold $\sqrt{Mm}$ has a competitive ratio
$\sqrt{\varphi}$, and any deterministic online algorithm attains a competitive ratio at least $\Omega(\sqrt{\varphi})$. 

In one-way (minimum) trading problem, a trader needs to exchange some currency to another currency. At each time slot $t$ the trader is presented a price $p(t)$ and must decide what portion of currency to be exchanged. \cite{Yaniv01} shows that the search problem and one-way trading problem are closely related: any deterministic (or randomized) one-way trading algorithm can be interpreted as a randomized search problem and vice versa. However, \cite{lorenz2009optimal} shows that randomization does not improve the competitive ratio of 1-min search problem (and one-way trading problem) better than $\Omega(\sqrt{\varphi})$. 

{\sf ESP} differs from 1-min search and one-trading problems in several aspects: (1) there are multiple demands appearing over time, (2) there is a limitation by the capacity of energy storage, (3) there are operational constraints of charging/discharging efficiencies and rates, and (4) there is also renewable energy.

\subsection{One-shot Decomposition}

\begin{definition}
Demand $(a(t))_{t=1}^{T}$ is called a {\em one-shot demand}, if there is a unique time slot ${t}_{\rm nz} \in [1, T]$ such that 
\begin{equation} 
a(t) = \left\{
\begin{array}{cl}
 0 & \mbox{if\ } t \ne t_{\rm nz} \\
 \bar{a} & \mbox{if\ } t = t_{\rm nz} \\
\end{array}
\right.
\end{equation}
where $\bar{a}$ is the peak demand of $(a(t))_{t=1}^{T}$. 
\end{definition}

\medskip

The simplest form of demand is ``one-shot demand''. 1-min search and one-trading problems can be regarded as a special case of {\sf ESP}, when there is zero amount of renewable energy (i.e., $\rho=0$), a one-shot demand with peak demand $\bar{a} = B$, and ideal energy storage (i.e., $\eta_{\rm c} = \eta_{\rm d} = 1$ and $\mu_{\rm c} = \mu_{\rm d} = \infty$). We next decompose demand $a(t)$ into a set of proper one-shot demands, and then solve each individually.

\medskip

\begin{definition}
Define a {\em one-shot decomposition} by:
\begin{equation}
{\sf 1sDecompose}\Big[(a(t))_{t=1}^{T}\Big]  =  (t_{\rm s}^i, t_{\rm nz}^i, \bar{a}^i )_{i = 1}^{m} 
\end{equation}
where $m$ is the number of decomposed one-shot demands, $t_{\rm nz}^i$ is the non-zero demand time slot and $\bar{a}^i$ is the peak demand of the $i$-th one-shot demand, and $t_{\rm s}^i$ $(\le t_{\rm nz}^i)$ is the minimum starting time slot for the $i$-th one-shot demand. ${\sf 1sDecompose}$ is subject to the following constraints:
\begin{enumerate}

\item[({\sf D1})] $a(t)$ can be reconstructed by the one-shot demands:
\begin{equation}
a(t) = \sum_{i:t_{\rm nz}^i = t} \bar{a}^i \mbox{\quad for all\ } t
\end{equation}

\item[({\sf D2})] There is a non-decreasing order on the starting time slots and non-zero demand time slots:
\begin{equation}
{t}_{\rm s}^i \le {t}_{\rm s}^{i+1} \mbox{\quad and\quad } {t}_{\rm nz}^i \le {t}_{\rm nz}^{i+1} \mbox{\quad for all\ } i
\end{equation}

\item[({\sf D3})] Let ${\cal D}$ be the set of one-shot demands that have non-zero duration, namely, ${\cal D} \triangleq \{i \mid {t}_{\rm s}^i < {t}_{\rm nz}^i \}$. Let ${\cal D}(i)$ be the set of one-shot demands in ${\cal D}$ other than $i$, such that the peak demands also lie within $[{t}_{\rm s}^i, {t}_{\rm nz}^i]$, namely, ${\cal D}(i) \triangleq \{j \in {\cal D} \backslash \{i\} \mid {t}_{\rm s}^i \le {t}_{\rm nz}^{j} \le {t}_{\rm nz}^i \}$. If $i \in {\cal D}$, then
\begin{equation} \label{eqn:constrictB}
\sum_{j: \in {\cal D}(i)} \bar{a}^{j} + \bar{a}^i \le \frac{B}{\eta_{\rm d}} 
\end{equation}
Eqn.~(\ref{eqn:constrictB}) ensures that satisfying the other one-shot demands in $[{t}_{\rm s}^i, {t}_{\rm nz}^i]$ using the energy storage still leaves sufficient capacity in the energy storage for the $i$-th one-shot demand. We set each ${t}_{\rm s}^i$ to be as minimum as possible subject to Eqn.~(\ref{eqn:constrictB}). When ${t}_{\rm s}^i = {t}_{\rm nz}^i$, then the one-shot demand needs to acquire energy from the grid.

\end{enumerate}
\end{definition}

\begin{figure}[htb!]
\hspace{-10pt}
\includegraphics[scale=0.4]{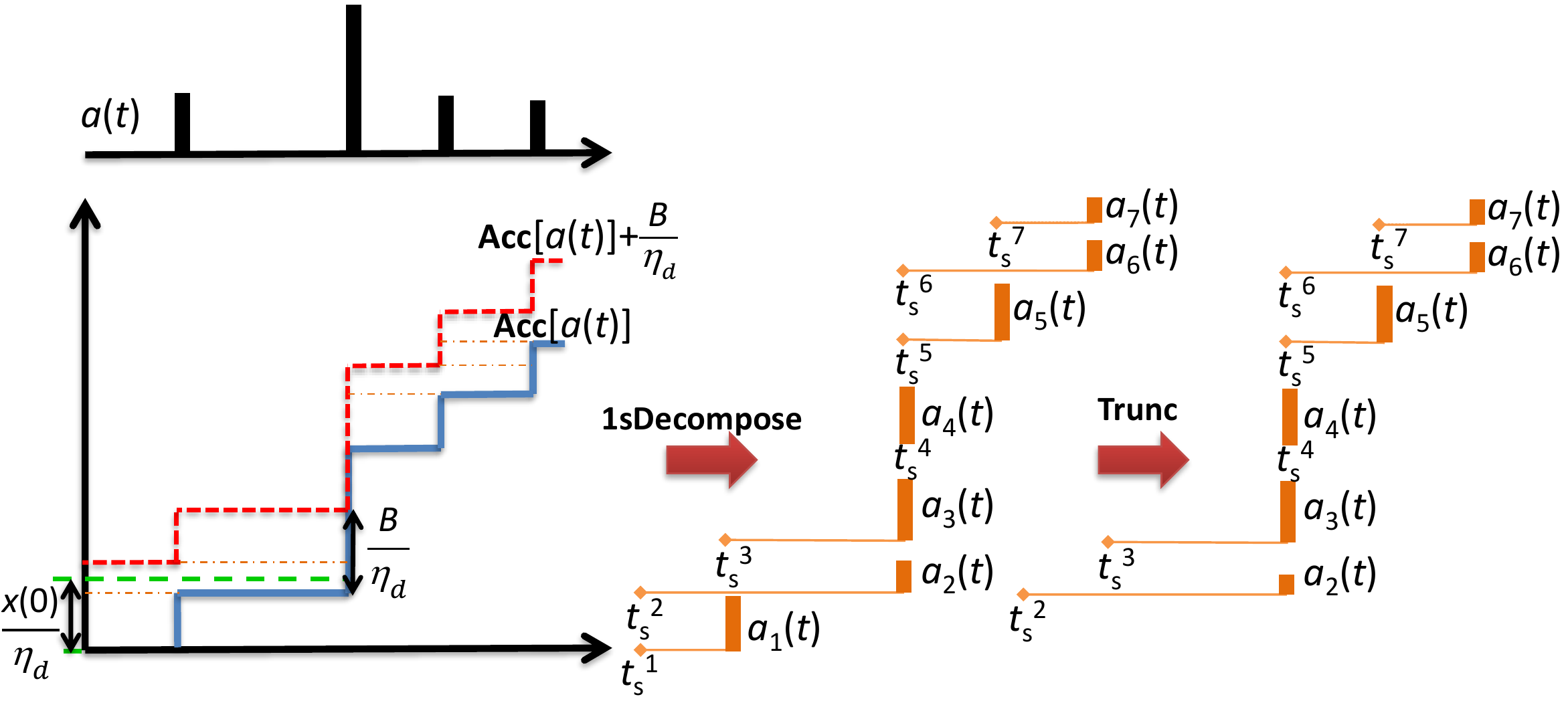}
\caption{An illustration of one-shot decomposition ${\sf 1sDecompose}$, which decomposes $a(t)$ into a set of one-shot demands that satisfy ({\sf D1})-({\sf D3}). The decomposed one-shot demands are truncated by ${\sf Trunc}[\cdot]$ to subtract those can be satisfied by the stored energy initially.} \label{fig:decompose}
\end{figure}

${\sf 1sDecompose}$ can be constructed in a simple manner. Define the accumulative demand curve by ${\sf Acc}[a(t)] =\sum_{t'=0}^{t} a(t')$, and ${\sf Acc}[a(t)]+ \frac{B}{\eta_{\rm d}} $ is the upward shift by $\frac{B}{\eta_{\rm d}} $. The one-shot demands are constructed by the rectanglizing the region sandwiched between ${\sf Acc}[a(t)]+ \frac{B}{\eta_{\rm d}} $ and ${\sf Acc}[a(t)]$. Each one-shot demand $(t_{\rm s}^i, t_{\rm nz}^i, \bar{a}^i )$ corresponds to a rectangle of $(t_{\rm nz}^i - t_{\rm s}^i)\times\bar{a}^i$, which is maximally inscribed in the sandwiched region. See Fig.~\ref{fig:decompose} for an illustration.

Since the initial level of energy storage $x(0)$ may be non-zero, we need to subtract the demand that can be satisfied by the stored energy initially.  
We define a function ${\sf Trunc}[\cdot]$, which truncates the decomposed one-shot demands that sum up to $\frac{x(0)}{\eta_{\rm d}}$ (according to the order output by ${\sf 1sDecompose}$). See Fig.~\ref{fig:decompose} for an illustration.

\subsection{Proof of Theorem~\ref{thm:online-comp3}}

\begin{customthm}{1} 
Suppose the terminal condition $x(T) = B$ and $\rho \le 1$.
Setting $\hat{B} = B(1-\rho)$ and
\begin{equation}
\theta = \frac{\sqrt{\rho^2 (M - m)^2 + 4 M m} - \rho ( M - m) }{2}  \cdot \frac{\eta_{\rm c}}{\eta_{\rm d}}, \label{eqn:threshold}
\end{equation}
the competitive ratio of online algorithm ${\cal A}_{\sf thb}$ is 
\begin{equation}
{\sf CR}({\cal A}_{\sf thb}) = \frac{1}{2} (\rho \varphi + \rho + \sqrt{ 4 \varphi + \rho^2 (\varphi - 1)^2})
\end{equation}
\end{customthm}

\begin{IEEEproof}
Let $(t_{\rm s}^i, t_{\rm nz}^i, \bar{a}^i )_{i = 1}^{m} = {\sf 1sDecompose}\big[(a(t))_{t=1}^{T}\big]$.

First, we consider $\widehat{\cal A}_{\sf ofl}$, which applies one-shot decomposition ${\sf 1sDecompose}$ and ${\sf Trunc}$, and then solves the each decomposed one-shot demand by ${\cal A}_{\sf ofl}$ sequentially. 
Since the one-shot decomposition satisfies ({\sf D1})-({\sf D3}), $\widehat{\cal A}_{\sf ofl}$ computes the offline optimal solution for $a(t)$, and 
 \begin{equation}
{\sf Cost}\big( \widehat{\cal A}_{\sf ofl}[a(t)]\big)
= \sum_{i = 1}^{m} {\sf Cost}\big( {\cal A}_{\sf ofl}[t_{\rm s}^i, t_{\rm nz}^i, \bar{a}^i]\big)
\end{equation}

\begin{algorithm}[htb!]
\caption{$\widehat{\cal A}_{\sf ofl}\big[\big(a(t),p(t),r(t)\big)_{t=1}^{T}\big]$}
\begin{algorithmic}[1]
\State $(t_{\rm s}^i, t_{\rm nz}^i, \bar{a}^i )_{i = 1}^{m} \leftarrow  {\sf Trunc}\big[{\sf 1sDecompose}\big[\big(a(t)\big)_{t=1}^{T}\big]\big]$
\For{each $i \in \{1,...,m\}$} 
\State $\big(x^i(t), d^i(t), r^i_{\rm b}(t), v^i_{\rm a}(t), v^i_{\rm b}(t)\big)_{t=t^i_{\rm s}}^{t^i_{\rm nz}}$
\Statex $\qquad \leftarrow{\cal A}_{\sf ofl}\big[x(t^i_{\rm s}-1), (t^i_{\rm s}, t^i_{\rm nz}, \bar{a}^i), \big(p(t),r(t)\big)_{t=t^i_{\rm s}}^{t^i_{\rm nz}} \big]$
\For{each $t \in [t^i_{\rm s}, t^i_{\rm nz}]$}	
\State $x(t) \leftarrow x(t) +  x^i(t)$,  $d(t) \leftarrow d(t) +  d^i(t)$
\State $r_{\rm b}(t) \leftarrow r_{\rm b}(t) +  r_{\rm b}^i(t)$, $v_{\rm a}(t) \leftarrow v_{\rm a}(t) +  v_{\rm a}^i(t)$
\State $v_{\rm b}(t) \leftarrow v_{\rm b}(t) +  v_{\rm b}^i(t)$
\LeftComment{Subtract consumed renewable energy}
\State $r(t) \leftarrow r(t) -  r_{\rm b}^i(t)$
\LeftComment{Subtract satisfied demand}
\State $a(t) \leftarrow a(t)-d^i(t)-v^i_{\rm a}(t)$
\EndFor 
\EndFor 
\State \Return $\big(x(t), d(t), r_{\rm b}(t), v_{\rm a}(t), v_{\rm b}(t)\big)_{t=1}^{T}$
\end{algorithmic}
\end{algorithm}

Next, define algorithm ${\cal A}_{\sf thb}^{\rm 1d}$ as ${\cal A}_{\sf thb}$, such that the input is one-shot demand $(t_{\rm s}, t_{\rm nz}, \bar{a})$, and $\hat{B} = \bar{a}(1-\rho)$. Let $\widehat{\cal A}_{\sf thb}$ be the algorithm which applies one-shot decomposition ${\sf 1sDecompose}$ and ${\sf Trunc}$, and then solves each decomposed one-shot demand by ${\cal A}_{\sf thb}^{\rm 1d}$ sequentially. Note that $\widehat{\cal A}_{\sf ofl}$ and $\widehat{\cal A}_{\sf thb}$ are introduced for the convenience of proof, as both have similar structures. 

It follows that
\begin{equation}
{\sf Cost}\big( \widehat{\cal A}_{\sf thb}[\theta, a(t)]\big)
= \sum_{i = 1}^{m} {\sf Cost}\big( {\cal A}_{\sf thb}^{\rm 1d}[\theta, t_{\rm s}^i, t_{\rm nz}^i, \bar{a}^i]\big)
\end{equation}

\begin{algorithm}
\caption{${\cal A}_{\sf thb}^{\rm 1d}[\theta, (t_{\rm s}, t_{\rm nz}, \bar{a}), p(t), r(t)]$}
\begin{algorithmic}[1]
\If{$t \ge t_{\rm s}$}
\State $\bar{a}(t) \leftarrow \left\{
\begin{array}{cl}
\bar{a} & \mbox{if\ } t = t_{\rm nz} \\
0 & \mbox{otherwise\ }
\end{array}
\right.$
\State \Return ${\cal A}_{\sf thb}[\theta, \bar{a} (1-\rho), t, \bar{a}(t), p(t), r(t)]$
\EndIf
\end{algorithmic}
\end{algorithm}

\begin{algorithm}[htb!]
\caption{$\widehat{\cal A}_{\sf thb}\big[\theta, \big(a(t),p(t),r(t)\big)_{t=1}^{T}\big]$}
\begin{algorithmic}[1]
\State $(t_{\rm s}^i, t_{\rm nz}^i, \bar{a}^i )_{i = 1}^{m} \leftarrow  {\sf Trunc}\big[{\sf 1sDecompose}\big[\big(a(t)\big)_{t=1}^{T}\big]\big]$
\For{each $t \in [t^i_{\rm s}, t^i_{\rm nz}]$}
\For{each $i \in \{1,...,m\}$} 
\If{$t \ge t_{\rm s}^i$ and $t \le t_{\rm nz}^i$}
\State $\big(x^i(t), d^i(t), r^i_{\rm b}(t), v^i_{\rm a}(t), v^i_{\rm b}(t)\big)$
\Statex $\qquad \qquad \quad \leftarrow{\cal A}_{\sf thb}^{\rm 1d}\big[\theta, (t^i_{\rm s}, t^i_{\rm nz}, \bar{a}^i),p(t),r(t) \big]$
\State $x(t) \leftarrow x(t) +  x^i(t)$,  $d(t) \leftarrow d(t) +  d^i(t)$
\State $r_{\rm b}(t) \leftarrow r_{\rm b}(t) +  r_{\rm b}^i(t)$, $v_{\rm a}(t) \leftarrow v_{\rm a}(t) +  v_{\rm a}^i(t)$
\State $v_{\rm b}(t) \leftarrow v_{\rm b}(t) +  v_{\rm b}^i(t)$, $r(t) \leftarrow r(t) -  r_{\rm b}^i(t)$
\State $a(t) \leftarrow a(t)-d^i(t)-v^i_{\rm a}(t)$
\EndIf
\EndFor 
\EndFor 
\State \Return $\big(x(t), d(t), r_{\rm b}(t), v_{\rm a}(t), v_{\rm b}(t)\big)_{t=1}^{T}$
\end{algorithmic}
\end{algorithm}

Note that the terminal condition $x(T) = B$ can be equivalently considered by adding a dummy demand $a(T) = \frac{B}{\eta_{\rm d} }$ in the end, and hence, fully charging the energy storage will not store unnecessary energy in the storage.

Since ${\cal A}_{\sf thb}$ and $\widehat{\cal A}_{\sf thb}$ use the same threshold $\theta$ and the one-shot decomposition satisfies ({\sf D1})-({\sf D3}), charging up the energy storage to level $\hat{B}(1-\rho)$ with respect to demand $a(t)$ in ${\cal A}_{\sf thb}$ is equivalent to charging up to level $\bar{a}(1-\rho)$ for each decomposed one-shot demand in $\widehat{\cal A}_{\sf thb}$. It follows that
\begin{equation}
{\sf Cost}\big( {\cal A}_{\sf thb}[\theta, a(t)]\big)
= \sum_{i = 1}^{m} {\sf Cost}\big( {\cal A}_{\sf thb}^{\rm 1d}[\theta, t_{\rm s}^i, t_{\rm nz}^i, \bar{a}^i]\big)
\end{equation}

Hence, the ratio of total costs can be decomposed as follows.
\begin{eqnarray} 
{\sf CR}({\cal A}_{\sf thb}) & = &
\max_{\sigma} \frac{{\sf Cost}\big( {\cal A}_{\sf thb}[a(t)]\big)}{{\sf Cost}\big( {\cal A}_{\sf ofl}[a(t)]\big)} \\
& = & \max_{\sigma} \frac{\sum_{i = 1}^{m} {\sf Cost}\big( {\cal A}_{\sf thb}^{\rm 1d}[t_{\rm s}^i, t_{\rm nz}^i, \bar{a}^i]\big)}{\sum_{i = 1}^{m} {\sf Cost}\big( {\cal A}_{\sf ofl}[t_{\rm s}^i, t_{\rm nz}^i, \bar{a}^i]\big)} \label{eqn:cr_sum} 
\end{eqnarray}

In the following, we first consider unconstrained charging and discharging rates, where $\mu_{\rm c}, \mu_{\rm d} \ge B$. 
With respect to each one-shot demand $(t_{\rm s}^i, t_{\rm nz}^i, \bar{a}^i)$, there are two cases:
\begin{enumerate}
\item[Case 1:]
Market price $p(t) > \theta$ for all $t \in [t_{\rm s}^i, t^i_{\rm nz}]$. Then ${\cal A}_{\sf thb}^{\rm 1d}$ will not store energy from the grid, but only from renewable energy. Let $\gamma^i \bar{a}^i$ be the amount of demand that can be satisfied by renewable energy stored in the energy storage, where $\gamma^i \le 1$. But ${\cal A}_{\sf thb}^{\rm 1d}$ needs to acquire energy from the grid for the rest of demand $(1-\gamma^i) \bar{a}^i$ at a market price at most $M$ at time slot $t^i_{\rm nz}$. ${\cal A}_{\sf ofl}$ needs to store energy from the grid for at least an amount of $(1-\gamma^i) \bar{a}^i \frac{\eta_{\rm d}}{\eta_{\rm c}}$ at a market price at least $\theta$ within $[t_{\rm s}^i, t^i_{\rm nz})$.
Hence,
\begin{equation*}
\begin{array}{@{}l@{\ }l@{}}
{\sf Cost}\big( {\cal A}_{\sf thb}^{\rm 1d}[t_{\rm s}^i, t_{\rm nz}^i, \bar{a}^i]\big) & \le (1-\gamma^i) \bar{a}^i M \notag \\
{\sf Cost}\big( {\cal A}_{\sf ofl}[t_{\rm s}^i, t_{\rm nz}^i, \bar{a}^i]\big) & \ge (1-\gamma^i)\bar{a}^i \theta\frac{\eta_{\rm d}}{\eta_{\rm c}}  \notag	
\end{array}
\end{equation*}

\item[Case 2:]
Market price $p(t) \le \theta$ for some $t \in [t_{\rm s}^i, t^i_{\rm nz}]$. Then ${\cal A}_{\sf thb}^{\rm 1d}$ will store energy from the grid for an amount $(1 - \rho) \bar{a}^i\frac{\eta_{\rm d}}{\eta_{\rm c}}$ at a market price at most $\theta$, and store the rest of energy from renewable energy, if available. However, if there is insufficient renewable energy, ${\cal A}_{\sf thb}^{\rm 1d}$ needs to acquire energy from the grid for an amount of $[\rho -\gamma^i]^+ \bar{a}^i$ at a market price at most $M$ at time slot $t^i_{\rm nz}$, whereas ${\cal A}_{\sf ofl}$ needs to acquire energy from the grid for an amount of $(1-\gamma^i) \bar{a}^i$ at a market price at least $m$.
Hence,
\begin{equation*}
\begin{array}{@{}l@{\ }l@{}}
{\sf Cost}\big( {\cal A}_{\sf thb}^{\rm 1d}[t_{\rm s}^i, t_{\rm nz}^i, \bar{a}^i]\big) & \le  (1 - \rho)\bar{a}^i\theta\frac{\eta_{\rm d}}{\eta_{\rm c}} + [\rho -\gamma^i]^+\bar{a}^i M \notag\\
{\sf Cost}\big( {\cal A}_{\sf ofl}[t_{\rm s}^i, t_{\rm nz}^i, \bar{a}^i]\big) & \ge (1-\gamma^i) \bar{a}^i m \notag
\end{array}
\end{equation*}
Note that $\gamma^i$ may be larger than $\rho$ for some $i$, but $\sum_{i=1}^m \gamma^i \bar{a}^i = \rho \sum_{i=1}^m  \bar{a}^i$. 

\end{enumerate}
Let $A_1$ be the set of indices of one-shot demands in Case 1, $A_2$ be the set of indices of one-shot demands in Case 2 such that $\gamma^i < \rho$, and $A_3$ be the set of indices of one-shot demands in Case 2 such that $\gamma^i \ge \rho$. Define ${\sf a} \triangleq \sum_{i=1}^m  \bar{a}^i$, and
\begin{equation*}
\begin{array}{@{}l@{\ }l@{\ }l@{\ }l@{}}
{\sf a}_1 \triangleq \sum_{i \in A_1} \bar{a}^i,& {\sf a}_2 \triangleq \sum_{i \in A_2} \bar{a}^i, & 
{\sf a}_3 \triangleq \sum_{i \in A_3} \bar{a}^i \\
{\sf b}_1 \triangleq \sum_{i \in A_1} \gamma^i  \bar{a}^i, & {\sf b}_2 \triangleq \sum_{i \in A_2} \gamma^i \bar{a}^i, & {\sf b}_3 \triangleq \sum_{i \in A_3} \gamma^i \bar{a}^i 
\end{array}
\end{equation*}
Note that ${\sf b}_1 \le {\sf a}_1$, ${\sf b}_2 \le \rho {\sf a}_2$, $\rho {\sf a}_3 \le {\sf b}_3 \le {\sf a}_3$, ${\sf a}_1 + {\sf a}_2 + {\sf a}_3 = {\sf a}$ and ${\sf b}_1 + {\sf b}_2 + {\sf b}_3 = \rho {\sf a}$.

Let 
\begin{equation*}
f({\sf a}_1, {\sf b}_1, {\sf a}_2, {\sf b}_2) \triangleq \textstyle  \frac{\zeta_1({\sf a}_1, {\sf b}_1) + \zeta_2({\sf a}_2, {\sf b}_2) + \zeta_3({\sf a}- {\sf a}_1- { \sf a}_2, \rho {\sf a}- {\sf b}_1- {\sf b}_2)}{\xi_1({\sf a}_1, {\sf b}_1) + \xi_2({\sf a}_2, {\sf b}_2) + \xi_3({\sf a}- {\sf a}_1- {\sf a}_2, \rho {\sf a}- {\sf b}_1- {\sf b}_2)},
\end{equation*}
where
\begin{equation*}
\begin{array}{@{}l@{\ }l@{}}
\zeta_1({\sf a}_1, {\sf b}_1) \triangleq ({\sf a}_1 \mbox{-} {\sf b}_1) M, & \xi_1({\sf a}_1, {\sf b}_1) \triangleq  ({\sf a}_1 \mbox{-} {\sf b}_1) \theta \frac{\eta_{\rm d}}{\eta_{\rm c}} \\
\zeta_2({\sf a}_2, {\sf b}_2) \triangleq (1 \mbox{-} \rho){\sf a}_2 \theta\frac{\eta_{\rm d}}{\eta_{\rm c}} + (\rho {\sf a}_2 \mbox{-} {\sf b}_2) M, & \xi_2({\sf a}_2, {\sf b}_2) \triangleq ({\sf a}_2 \mbox{-} {\sf b}_2) m \\
\zeta_3({\sf a}_3, {\sf b}_3) \triangleq (1 \mbox{-} \rho){\sf a}_3 \theta\frac{\eta_{\rm d}}{\eta_{\rm c}}, & \xi_3({\sf a}_3, {\sf b}_3) \triangleq ({\sf a}_3 \mbox{-} {\sf b}_3) m 
\end{array}
\end{equation*} 
Let ${\cal F}$ be the feasible set of $({\sf a}_1, {\sf b}_1, {\sf a}_2, {\sf b}_2)$ subject to ${\sf b}_1 \le {\sf a}_1$, ${\sf b}_2 \le \rho {\sf a}_2$ and $\rho ({\sf a} - {\sf a}_1 - {\sf a}_2) \le \rho {\sf a} - {\sf b}_1 - {\sf b}_2 \le {\sf a} - {\sf a}_1 - {\sf a}_2$.

The competitive ratio from Eqn.~(\ref{eqn:cr_sum}) is obtained as follows.
\begin{eqnarray}
{\sf CR}({\cal A}_{\sf thb}) & = & \displaystyle \max_{({\sf a}_1, {\sf b}_1, {\sf a}_2, {\sf b}_2) \in {\cal F}} f({\sf a}_1, {\sf b}_1, {\sf a}_2, {\sf b}_2)\\
& \le & \max\big\{
\frac{\theta \frac{\eta_{\rm d}}{\eta_{\rm c}} + \rho M }{ m }, 
\frac{M + \rho \theta  \frac{\eta_{\rm d}}{\eta_{\rm c}} }{\theta  \frac{\eta_{\rm d}}{\eta_{\rm c}}} \big\}  \label{eqn:ub-cr}
\end{eqnarray}
where Eqn.~(\ref{eqn:ub-cr}) follows from Lemma~\ref{lem:fmax}. 

By substituting $({\sf a}_1 = {\sf b}_1 = 0, {\sf a}_2 = (1-\rho) {\sf a}, {\sf b}_2 = 0, {\sf a}_3 = {\sf b}_3 = \rho {\sf a})$, we obtain $f({\sf a}_1, {\sf b}_1, {\sf a}_2, {\sf b}_2) = \frac{\theta \frac{\eta_{\rm d}}{\eta_{\rm c}} + \rho M }{ m }$. Also, by substituting $({\sf a}_1 = (1-\rho) {\sf a}, {\sf b}_1 = 0, {\sf a}_2 = {\sf b}_2 = 0, {\sf a}_3 = {\sf b}_3 = \rho{\sf a})$, we obtain $f({\sf a}_1, {\sf b}_1, {\sf a}_2, {\sf b}_2) = \frac{M + \rho \theta  \frac{\eta_{\rm d}}{\eta_{\rm c}} }{\theta  \frac{\eta_{\rm d}}{\eta_{\rm c}}}$. Hence, 
${\sf CR}({\cal A}_{\sf thb}) =
\max\big\{
\frac{\theta \frac{\eta_{\rm d}}{\eta_{\rm c}} + \rho M }{ m }, 
\frac{M + \rho \theta  \frac{\eta_{\rm d}}{\eta_{\rm c}} }{\theta  \frac{\eta_{\rm d}}{\eta_{\rm c}}} \big\}$.

An adversary will select the worst among the two options in Eqn.~(\ref{eqn:ub-cr}). In order to minimize the competitive ratio, we set $\frac{\theta \frac{\eta_{\rm d}}{\eta_{\rm c}} + \rho M }{ m } = 
\frac{M + \rho \theta  \frac{\eta_{\rm d}}{\eta_{\rm c}} }{\theta  \frac{\eta_{\rm d}}{\eta_{\rm c}}}$.
Equivalently, $(\theta  \frac{\eta_{\rm d}}{\eta_{\rm c}})^2 + \rho (M - m) (\theta  \frac{\eta_{\rm d}}{\eta_{\rm c}}) - M m = 0$,
where $\theta = \frac{\sqrt{\rho^2 (M - m)^2 + 4 M m} - \rho ( M - m) }{2}  \frac{\eta_{\rm c}}{\eta_{\rm d}}$ is the positive root. Thus, the competitive ratio is
\begin{equation} 
{\sf CR}({\cal A}_{\sf thb}) =
\frac{\theta \frac{\eta_{\rm d}}{\eta_{\rm c}} + \rho M }{m}  = 
\frac{1}{2} (\rho \varphi + \rho + \sqrt{ 4 \varphi + \rho^2 (\varphi - 1)^2}) \notag
\end{equation}

Finally, we consider constrained charging and discharging rates, where $\mu_{\rm c} < B$ or $\mu_{\rm d} < B$. If $\mu_{\rm c} < B$, then there exists a decomposed one-shot demand that needs to store energy from the grid in at least two time slots, instead of one time slot with unconstrained charging rate. In this case, the offline optimal solution may not always be able to acquire energy from the grid at the time slot of the lowest market price. Threshold-based ${\cal A}_{\sf thb}$ uses the same threshold $\theta$ throughout the process. Therefore, the competitive ratio with constrained charging rate is not higher than the one with unconstrained charging rate.

If $\mu_{\rm d} < B$, then some demand may not be satisfied by energy storage as in the setting of unconstrained discharging, but by acquiring energy from the grid at the moment the demand arrives. This affects both online and offline algorithms to the same extent. Therefore, the competitive ratio with constrained discharging rate is not higher than the one with unconstrained discharging rate.
\end{IEEEproof}

\medskip

\begin{lemma} \label{lem:fmax}
\begin{equation*} \hspace{-10pt}
\max_{({\sf a}_1, {\sf b}_1, {\sf a}_2, {\sf b}_2) \in {\cal F}} f({\sf a}_1, {\sf b}_1, {\sf a}_2, {\sf b}_2) \le \max\big\{
\frac{\theta \frac{\eta_{\rm d}}{\eta_{\rm c}} + \rho M }{ m }, 
\frac{M + \rho \theta  \frac{\eta_{\rm d}}{\eta_{\rm c}} }{\theta  \frac{\eta_{\rm d}}{\eta_{\rm c}}} \big\}
\end{equation*}
where $m \le \theta \frac{\eta_{\rm d}}{\eta_{\rm c}} \le \sqrt{M m}$.
\end{lemma}
\begin{IEEEproof}
The maximum of $f({\sf a}_1, {\sf b}_1, {\sf a}_2, {\sf b}_2)$ is attained either as an interior point (i.e., ${\sf a}_1 > 0, {\sf a}_2 > 0, {\sf a}_3 > 0$) or a boundary point (i.e., ${\sf a}_1 = 0$, or ${\sf a}_2 = 0$, or ${\sf a}_3 = 0$). 
Suppose that the maximum is an interior point, then it is a stationary point, namely, the partial derivative of $f({\sf a}_1, {\sf b}_1, {\sf a}_2, {\sf b}_2)$ at each parameter is zero. However, by differentiation, one obtains
\begin{equation*}
\begin{array}{@{}l@{\ }l@{}}
\frac{\partial f({\sf a}_1, {\sf b}_1, {\sf a}_2, {\sf b}_2)}{\partial {\sf a}_2} = \frac{M \rho}{(\theta\frac{\eta_{\rm d}}{\eta_{\rm c}} - m )({\sf a}_1 - {\sf b}_1) + {\sf a} m (1- \rho)} \ne 0
\end{array}
\end{equation*}
Hence, the maximum of  $f({\sf a}_1, {\sf b}_1, {\sf a}_2, {\sf b}_2)$ is a boundary point. We consider a boundary point by three cases:
\begin{enumerate}

\item ${\sf a}_1 = 0$: This implies that ${\sf b}_1 = 0$ because $0 \le {\sf b}_1 \le {\sf a}_1$. \hspace{-20pt}
\begin{equation*}
\begin{array}{@{}l@{\ }l@{}}
 f(0, 0, {\sf a}_2, {\sf b}_2) =  \frac{\zeta_2({\sf a}_2, {\sf b}_2) + \zeta_3({\sf a}\mbox{-}{ \sf a}_2, \rho {\sf a}\mbox{-}{\sf b}_2)}{\xi_2({\sf a}_2, {\sf b}_2) + \xi_3({\sf a}\mbox{-}{\sf a}_2, \rho {\sf a}\mbox{-}{\sf b}_2)} \stackrel{(a)}{\le} 
 \frac{\theta \frac{\eta_{\rm d}}{\eta_{\rm c}} + \rho M }{ m }
\end{array}
\end{equation*}
Inequality (a) follows from algebraic operations:
\begin{equation*}
\begin{array}{@{}l@{\ }l@{}}
&  \big(\xi_2({\sf a}_2, {\sf b}_2) + \xi_3({\sf a}-{\sf a}_2, \rho {\sf a}-{\sf b}_2)\big) \big(\theta \frac{\eta_{\rm d}}{\eta_{\rm c}} + \rho M\big) \notag \\ 
&  - \big(\zeta_2({\sf a}_2, {\sf b}_2) + \zeta_3({\sf a}- { \sf a}_2, \rho {\sf a}\mbox{-}{\sf b}_2) \big) (m) \notag \\
= & m M \big( \rho {\sf a} - \rho {\sf a}_2 - (\rho^2 {\sf a} - {\sf b}_2) \big) \ge m M \rho( {\sf a}_3 - {\sf b}_3)  \ge 0  \notag
\end{array}
\end{equation*}

\item ${\sf a}_2 = 0$: Also, ${\sf b}_2 = 0$ because $0 \le {\sf b}_2 \le {\sf a}_2$.  \hspace{-20pt}
\begin{equation*}
\begin{array}{@{}l@{\ }l@{}}
 f({\sf a}_1, {\sf b}_1, 0, 0) & =  \frac{\zeta_1({\sf a}_1, {\sf b}_1) + \zeta_3({\sf a}\mbox{-}{ \sf a}_1, \rho {\sf a} \mbox{-}{\sf b}_1)}{\xi_1({\sf a}_1, {\sf b}_1) + \xi_3({\sf a}\mbox{-}{\sf a}_1, \rho {\sf a}\mbox{-}{\sf b}_1)} \stackrel{(b)}{\le} \frac{M + \rho \theta  \frac{\eta_{\rm d}}{\eta_{\rm c}} }{\theta  \frac{\eta_{\rm d}}{\eta_{\rm c}}}
\end{array}
\end{equation*}
Inequality (b) follows from algebraic operations:
\begin{equation*}
\begin{array}{@{}l@{\ }l@{}}
&  \big(\xi_1({\sf a}_1, {\sf b}_1) + \xi_3({\sf a}- {\sf a}_1, \rho {\sf a}-{\sf b}_1)\big) \big(M + \rho \theta  \frac{\eta_{\rm d}}{\eta_{\rm c}}\big) \notag \\ 
&  - \big(\zeta_1({\sf a}_1, {\sf b}_1) + \zeta_3({\sf a}- { \sf a}_1, \rho {\sf a} - {\sf b}_1) \big) \big( \theta  \frac{\eta_{\rm d}}{\eta_{\rm c}} \big) \notag \\
 = & {\sf b}_1 \big( m (M + \rho \theta  \frac{\eta_{\rm d}}{\eta_{\rm c}}) - \rho \big(\theta  \frac{\eta_{\rm d}}{\eta_{\rm c}} \big)^2\big) \notag \\
 & - ({\sf a}_1 - (1-\rho) {\sf a}) \big( m (M + \rho \theta  \frac{\eta_{\rm d}}{\eta_{\rm c}}) - \big(\theta  \frac{\eta_{\rm d}}{\eta_{\rm c}} \big)^2\big) \ge 0 \notag
\end{array}
\end{equation*}
Because $ {\sf b}_1 \ge {\sf a}_1 - (1-\rho) {\sf a} \Leftrightarrow {\sf a}_3 \ge {\sf b}_3$, and $\theta \frac{\eta_{\rm d}}{\eta_{\rm c}} \le \sqrt{M m}$.

\item ${\sf a}_3 = 0$: Also, ${\sf b}_3 = 0$ because $0 \le {\sf b}_3 \le {\sf a}_3$. Namely, ${\sf a}_2 = {\sf a} - {\sf a}_1$ and ${\sf b}_2 = \rho {\sf a} - {\sf b}_1$. By algebraic operations, one can show that  \hspace{-20pt}
\begin{equation*}
\begin{array}{@{}l@{\ }l@{}}
 f({\sf a}_1, {\sf b}_1, {\sf a}\mbox{-}{\sf a}_1, \rho{\sf a}\mbox{-}{\sf b}_1) 
 & \le \max\big\{
\frac{\theta \frac{\eta_{\rm d}}{\eta_{\rm c}} }{ m }, 
\frac{M }{\theta  \frac{\eta_{\rm d}}{\eta_{\rm c}}} \big\} \\
 & \le 
\max\big\{
\frac{\theta \frac{\eta_{\rm d}}{\eta_{\rm c}} + \rho M }{ m }, 
\frac{M + \rho \theta  \frac{\eta_{\rm d}}{\eta_{\rm c}} }{\theta  \frac{\eta_{\rm d}}{\eta_{\rm c}}} \big\} \notag
\end{array}
\end{equation*}
\end{enumerate}
Therefore, the upper bound of $f({\sf a}_1, {\sf b}_1, {\sf a}_2, {\sf b}_2)$ is the maximum of the three cases of boundary points.
\end{IEEEproof}

\subsection{Proof of Theorem~\ref{thm:lb-det0}}

\begin{customthm}{2}  
Consider zero excessive renewable energy ($\rho=0$). The competitive ratio of any deterministic online algorithm ${\cal A}$ is ${\sf CR}({\cal A}) \ge \frac{1}{2}\big(1+\sqrt{\varphi}\big)$.
\end{customthm}

\begin{IEEEproof}
It suffices to show that it is true for certain inputs. Consider $\mu_{\rm c}, \mu_{\rm d} \ge B$, boundary conditions $x(0)=x(T)=B$, and the two following inputs with $T =3$:
\begin{enumerate}

\item Input $\sigma_1 = \big( a_1(t), p_1(t) \big)_{t=1}^3$: 
\begin{equation}
\begin{array}{@{}l@{\ }l@{\ }l@{}}
a_1(1)=\frac{B}{\eta_{\rm d}}, & a_1(2)=0, & a_1(3)=0 \\
p_1(1)=M, & p_1(2)=\sqrt{Mm}, & p_1(3)=m
\end{array}
\end{equation}

\item Input $\sigma_2 = \big( a_2(t), p_2(t) \big)_{t=1}^3$: 
\begin{equation}
\begin{array}{@{}l@{\ }l@{\ }l@{}}
a_2(1)=\frac{B}{\eta_{\rm d}}, & a_2(2)=0, & a_2(3)=0 \\
p_2(1)=M, & p_2(2)=\sqrt{Mm}, & p_2(3)=M
\end{array}
\end{equation}

\end{enumerate}

Note that any feasible deterministic online algorithm ${\cal A}$ will discharge the energy storage from $B$ to $0$ for $\sigma_1$ and $\sigma_2$ to satisfy the demands $a_1(1)$ and $a_2(1)$ because the market prices $p_1(1)$ and $p_2(1)$ attain the maximum value $M$.

Then at $t=2$, both $\sigma_1$ and $\sigma_2$ are indistinguishable. Let the amount of energy charged to the energy storage by ${\cal A}$ at $t=2$ be $z$. Both online and offline must store energy up to level $B$ from the grid because of the terminal condition $x(T)=B$.
\begin{enumerate}

\item
For $\sigma_1$, ${\sf Opt}(\sigma_1) = m\frac{B}{\eta_{\rm c}}$
and 
\begin{equation}
{\sf Cost}({\cal A}[\sigma_1]) = \frac{z}{\eta_{\rm c}}\sqrt{Mm}+m\frac{B-z}{\eta_{\rm c}}
\end{equation}

\item
For $\sigma_2$, ${\sf Opt}(\sigma_2) =\sqrt{Mm} \frac{B}{\eta_{\rm c}}$
and 
\begin{equation}
{\sf Cost}({\cal A}[\sigma_2]) = \frac{z}{\eta_{\rm c}}\sqrt{Mm}+M\frac{B-z}{\eta_{\rm c}}
\end{equation}

\end{enumerate}
The adversary can choose either $\sigma_1$ or $\sigma_2$ to maximize the competitive ratio, whereas ${\cal A}$ needs to optimize $z$ to minimize the competitive ratio.
Hence, the competitive ratio of ${\cal A}$ is lower bounded by:
\begin{equation} \hspace{-10pt}
\begin{array}{@{}r@{\ }r@{\ }l@{}}
{\sf CR}({\cal A}) & \ge & \min_{z}\Big( \max\Big\{\frac{\frac{z}{\eta_{\rm c}}\sqrt{Mm}+m\frac{B-z}{\eta_{\rm c}}}{m\frac{B}{\eta_{\rm c}}},\frac{\frac{z}{\eta_{\rm c}}\sqrt{Mm}+M\frac{B-z}{\eta_{\rm c}}}{\frac{B}{\eta_{\rm c}}\sqrt{Mm}}\Big\} \Big)\\
 & = & \min_{z}\Big( \max\Big\{\frac{z\sqrt{Mm}+m(B-z)}{m},\frac{z\sqrt{Mm}+M(B-z)}{\sqrt{Mm}}\Big\} \Big)\\
 & = & \frac{1}{2}\big(1+\sqrt{\varphi}\big)
\end{array}
\end{equation}
where the minimum is attained at value $z=\frac{B}{2}$, when
$\frac{z\sqrt{Mm}+m(B-z)}{m}=\frac{z\sqrt{Mm}+M(B-z)}{\sqrt{Mm}}$.
\end{IEEEproof}

\subsection{Proof of Theorem~\ref{thm:lb-det1}}

\begin{customthm}{3} 
Consider abundant excessive renewable energy ($\rho\ge1$). The competitive ratio of any deterministic online algorithm ${\cal A}$ is ${\sf CR}({\cal A}) \ge \varphi$.
\end{customthm}

\begin{IEEEproof}
Given deterministic online algorithm ${\cal A}$, we input $p(t)=m, a(t)=0, r(t)=0$ at every time slot $t$ until it is observed that ${\cal A}$ stores energy from the grid, say at time slot $t'$. Then, we input $r(t'+1)=\frac{B}{\eta_{\rm c}}, p(t'+2)=M, a(t'+2)= \frac{B}{\eta_{\rm d}}$ (which is always possible because $\rho\ge1$). If ${\cal A}$ ever stores energy from the grid, then the competitive ratio is unbounded, as the offline optimal cost is zero (from free renewable energy). If ${\cal A}$ has a bounded competitive ratio, then ${\cal A}$ never stores energy from the grid. In this case, the ${\sf CR}({\cal A}) = \varphi$.
\end{IEEEproof}

\end{document}